\documentclass[reprint,superscriptaddress,amsmath,amssymb,aps,prx]{revtex4-2}
\usepackage[scaled]{beramono}
\usepackage[T1]{fontenc}
\usepackage{graphicx, amsfonts, amsmath, amssymb, cancel, bm}
\usepackage[colorlinks=true]{hyperref}
\usepackage{siunitx}
\usepackage{xcolor}
\usepackage{colortbl}
\usepackage{commath}
\usepackage{bbm}

\usepackage[capitalize]{cleveref}

\let\Re\undefined
\let\Im\undefined
\DeclareMathOperator{\Re}{{\rm Re}}
\DeclareMathOperator{\Im}{{\rm Im}}

\newcommand{\cit}[1]{Ref.~\cite{#1}}
\newcommand*\diff{\mathop{}\!\mathrm{d}}

\newcommand{\nn}{\nonumber\\}

\newcommand{\me}{\mathrm{e}}
\newcommand{\im}{\mathrm{i}\mkern1mu}

\begin{document}

\title{Equivalent circuit and continuum modeling of the impedance of electrolyte-filled pores}

\author{Christian Pedersen}
\email{chrpe@math.uio.no}
\affiliation{Department of Mathematics, Mechanics Division, University of Oslo, N-0851 Oslo, Norway}

\author{Timur Aslyamov}
\email{timur.aslyamov@uni.lu}
\affiliation{Complex Systems and Statistical Mechanics, Department of Physics and Materials Science, University of Luxembourg, L-1511 Luxembourg City, Luxembourg}

\author{Mathijs Janssen}
\email{mathijs.a.janssen@nmbu.no}
\affiliation{Norwegian University of Life Sciences, Faculty of Science and Technology, Pb 5003, 1433 Ås, Norway}

\date{\today}

\begin{abstract}
Batteries, supercapacitors, and several other electrochemical devices  charge by accumulating ions in the pores of electrolyte-immersed porous electrodes. 
The charging of such devices has long been interpreted using equivalent circuits and the partial differential equations these give rise to. 
Here, we discuss the validity of the transmission line (TL) circuit and equation for modeling a single electrolyte-filled pore in contact with a reservoir of resistance $R_{r}$.
The textbook derivation of the pore-reservoir impedance $R_r+Z_p$ from the TL equation does not correctly account for ionic current conservation at the pore-reservoir interface.
However, correcting this shortcoming leads to the same impedance.
We also show that the pore impedance $Z_p$ can be derived directly from the TL circuit,  bypassing the TL equation completely.
The TL circuit assumes equipotential lines in an electrolyte-filled pore to be straight, which is not the case near the pore entrance and end.
To determine the importance of these regions, we numerically simulated the charging of pores of different lengths $\ell_p$ and radii $\varrho_p$ through the Poisson-Nernst-Planck equations.
We find that pores with aspect ratios beyond $\ell_p/\varrho_p\gtrapprox5$ have impedances in good agreement with $Z_p$.
\end{abstract}

\maketitle

\section{Introduction}
\subsection{The physics of charging porous electrodes}
\begin{figure}
    \includegraphics[width=\linewidth]{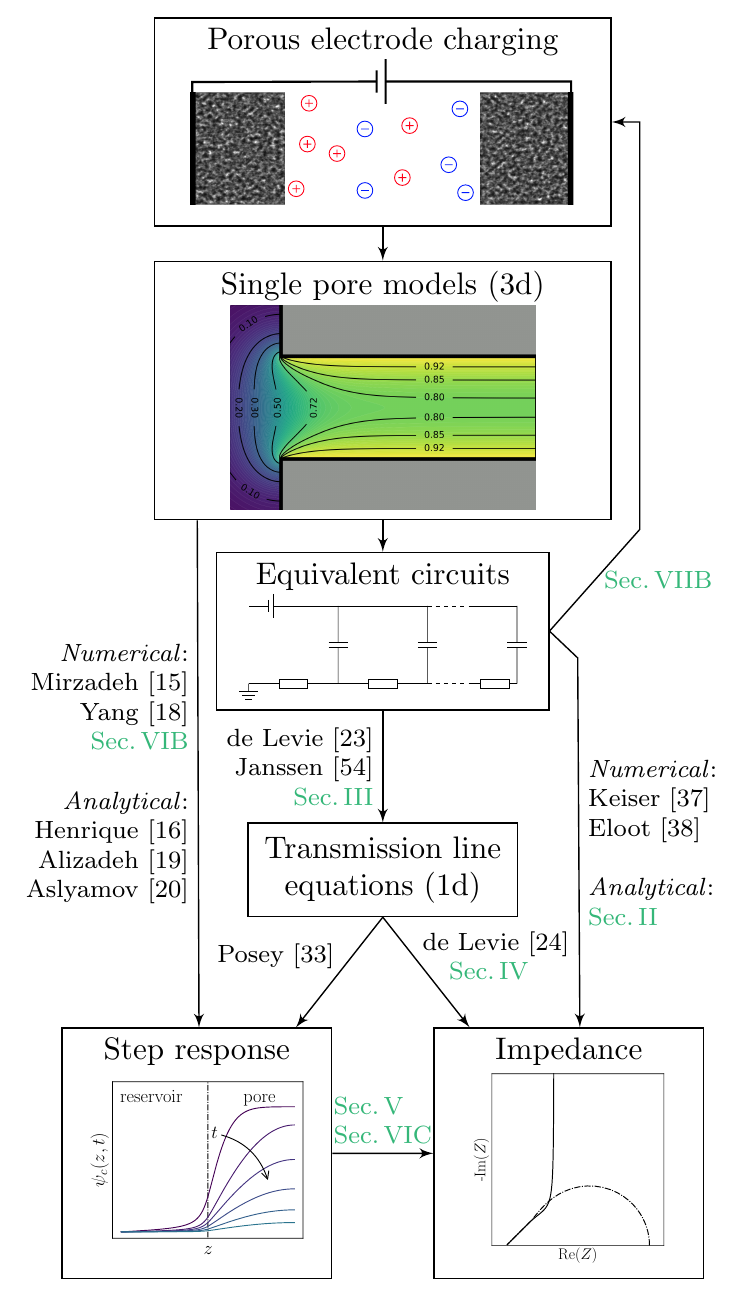}
    \caption{
    Overview of modeling approaches to understand the response of porous electrodes to an applied potential.
    The figure mentions a few representative references by the first authors' names; see the introduction for more references and the bibliography for full information. 
    The new contributions of this article are indicated in green. 
    \label{fig0}}
\end{figure}

Electrolyte-immersed porous electrodes are used in several technologies, including in batteries \cite{moskon2021transmission}, solid oxide fuel cells \cite{nielsen2014impedance}, electrochemical sensors \cite{privett2010electrochemical}, supercapacitors \cite{kotz2000principles,wu2022understanding}, and capacitive deionization devices \cite{porada2013review}. 
In these applications, the porous electrodes typically contain pores of different shapes, widths, and lengths, connected hierarchically.
When a potential difference is applied between two porous electrodes, migration of ions in electric fields leads, in each electrode, to the accumulation of one type of ion and an opposing electric charge on the electrode surface, which together are called the electric double layer (EDL) (see \cref{fig0} for a schematic summary of the Introduction).
Ions also diffuse if they pile up or dwindle locally and convect if the applied potential drives electro-osmosis \cite{ratschow2022resonant}.
Lastly, narrow pores can contain only so many finite-size ions, so the ionic fluxes are also affected by steric repulsions \cite{kilic2007stericII,aslyamov2022relation,tomlin2022impedance}. 
A theoretical model for all these effects should involve at least the Poisson equation for the electrostatics, the Navier-Stokes equation for the fluid flow, and modified Nernst-Planck equations to describe the flux of finite-size ions; solvent-free ionic liquids would require a yet-to-be-developed continuum model instead \cite{lee2015dynamics}.
These equations should be solved in a porous electrode's 3d geometry, resolving charge storage in nanometre-wide pores and ionic fluxes through mesopores and between the electrodes over micrometers.
Currently, computational resources do not allow one to do so.
Many models for porous electrode charging thus ignore their large-scale structure and instead focus on the charging of idealized pores, usually either a few nanometres or micrometers wide (see the second box in \cref{fig0}).
Fluid flow is also often neglected, which is apposite for small applied potentials \cite{malgaretti2019driving}. 
The resulting Poisson-Nernst-Planck (PNP) equations were solved numerically \cite{sakaguchi2007,lim2009effect,mirzadeh2014enhanced,henrique2021charging,henrique2022impact,yang2022simulating} and analytically \cite{alizadeh2017multiscale, henrique2021charging,henrique2022impact,aslyamov2022analytical}.
While single-pore models oversimplify the charging of a porous electrode, numerically solving the PNP equations in a single pore is still computationally expensive, so the first mentioned studies go back less than two decades.

\subsection{Single-pore equivalent circuit models}
Historically, porous electrode charging was first studied through circuit models \cite{danielbek1948,ksenzhek1956,levie1963}.
Again, rather than an entire porous electrode, these works considered the charging of a single pore.
The electrolyte in a pore has a resistance ($R_p$) and the electrolyte-electrode interface has a capacitance ($C$), but a pore does not charge like an $RC$ circuit because the resistance and capacitance are distributed over the pore, which can be represented by cutting up $R_p$ and $C$ connecting the pieces in the transmission line (TL) circuit---the ladder network shown in the third box in \cref{fig0}.
In the limit of infinitely many, infinitesimally small circuit elements, the TL circuit gives rise to the TL equation [viz. \cref{eq:TLeqfaradaic}], a diffusion-type equation for the potential drop across the capacitors of the circuit. 
De Levie solved the TL equation for a case of a finite-length pore of constant cross-section and capacitance subject to a sinusoidal applied voltage of angular frequency $\omega $, yielding the pore impedance \cite{levie1967electrochemical},
\begin{equation}\label{eq:ZWO}
    Z_p =\sqrt{\frac{R_p }{\im\omega C} } \coth \sqrt{ \im\omega R_p C },
\end{equation}
where $\im=\sqrt{-1}$. 
The mathematical form $\coth{\sqrt{(.)}}/\sqrt{(.)}$ is typical for diffusion in bounded geometries---it also arises for finite-length mass transfer of electroactive species to a planar electrode, where it is called the Warburg open impedance \cite{orazem2008electrochemical,lasia2014electrochemical}.

\Cref{eq:ZWO} has been widely used to interpret electrochemical impedance spectroscopy (EIS) experiments on porous electrodes, often in combination with other circuit elements \cite{gassa1990electrochemical,jurczakowski2004impedance,ogihara2012theoretical,ogihara2015impedance}.
For instance, the impedance of a pore in contact with an electrolyte reservoir of resistance $R_{r}$ reads 
\begin{equation}\label{eq:ZWO_total}
    Z=R_{r}+\sqrt{\frac{R_p }{\im\omega C} } \coth \sqrt{ \im\omega R_p C }.
\end{equation}
When viewing $Z_p$ and $R_{r}$ as circuit elements, \cref{eq:ZWO_total} follows from \cref{eq:ZWO} as the impedances of circuit elements in series can be simply added. 
In terms of the underlying physics, however, adding these separate pore and reservoir impedances makes less sense.
De Levie's derivation of \cref{eq:ZWO} employed a boundary condition corresponding to a counter electrode placed at the pore entrance, effectively setting the reservoir's resistance to zero.
So \cref{eq:ZWO_total} reintroduces the reservoir resistance after first setting it to zero.
This procedure does not correctly account for ionic flux conservation at the pore-reservoir interface [see \cref{sec:Zreflectivebc}].
Still, de Levie's derivation of \cref{eq:ZWO} is repeated unaltered in recent textbooks and reviews \cite{conway2013electrochemical,lasia2014electrochemical,huang2020editors}.
Shortly after de Levie \cite{levie1963,levie1967electrochemical}, Posey and Morozumi used the correct boundary condition in their study of the TL model's step response \cite{posey1966theory}.
One of the contributions of this article is that we show that \cref{eq:ZWO_total} also follows from the TL equation using Posey and Morozumi's correct boundary condition.

\begin{figure}
    \includegraphics[width=\linewidth]{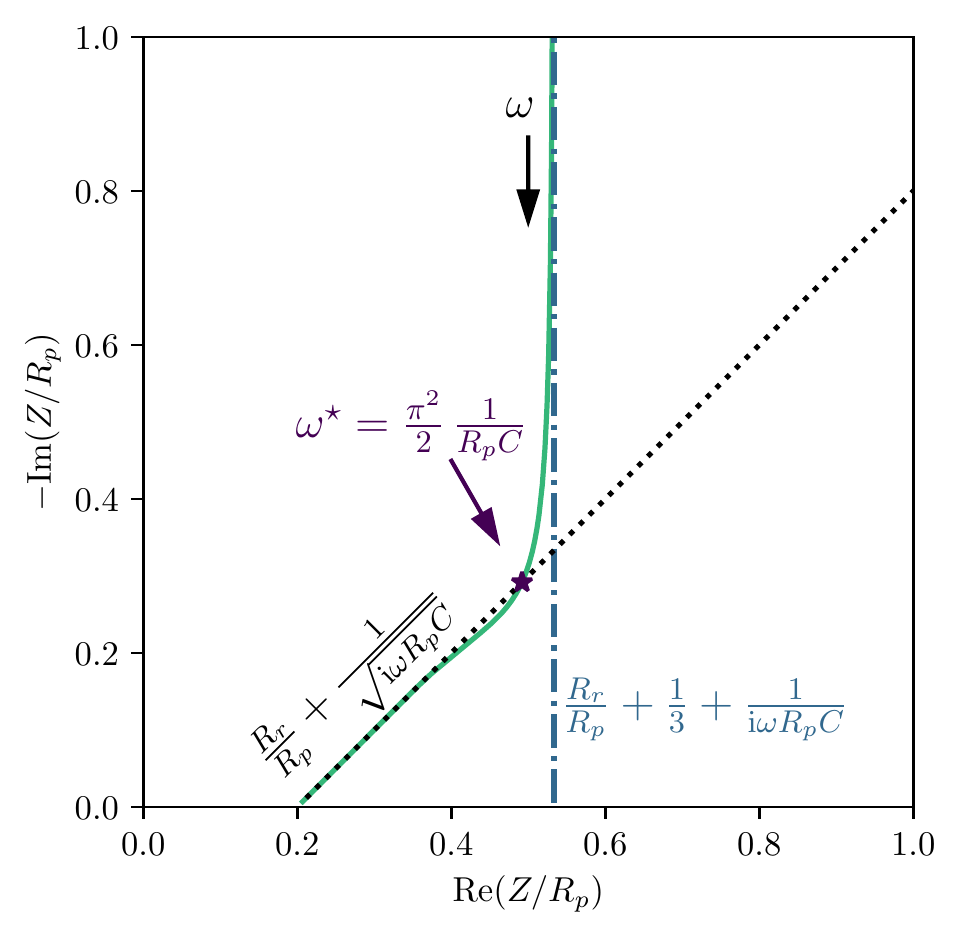}
    \caption{Complex plane plot of \cref{eq:ZWO_total} for $R_{r}/R_p=0.2$. \label{fig_Warburg}}
\end{figure}

\Cref{fig_Warburg} is a ``complex plane plot'' of \cref{eq:ZWO_total}, showing its real versus its imaginary part for different $\omega$.
The plot shows a 45-degree line at high frequencies, characteristic of diffusion in semi-infinite geometries, and a 90-degree line at low frequencies, characteristic of capacitor charging.
The transition between these two regimes occurs around the frequency $\omega^{\star} =\pi^2/(2R_p C )$ \cite{janssen2021locating}, and the other indicated formulas follow from the limits $Z_p(\omega\to\infty)=\sqrt{R_p/(\im\omega C)}$ and $Z_p(\omega\to0)=R_p/3+ 1/\im\omega C$. 

\Cref{eq:ZWO,eq:ZWO_total} apply to a case where all the elements in the TL ladder circuit have the same resistance and capacitance; that is, the resistance and capacitance are constant along the pore.
Hence, for a pore of length $\ell_p$, surface area $A^s_p$, and arbitrary but fixed cross-sectional area $A^c_p$, in contact with a reservoir of length $\ell_r$ and fixed cross-sectional area $A^c_r$, we have
\begin{equation}\label{eq:resistanceRandcapacitancC}
    R_p=\frac{1}{\kappa}\frac{\ell_p}{A^c_p},\qquad R_{r}=\frac{1}{\kappa}\frac{\ell_r}{A^c_r},\qquad C=c_{\rm EDL}A^s_p,
\end{equation}
where $\kappa$ is the electrolyte conductivity and $c_{\rm EDL}$ is the EDL capacitance per unit electrode area.
To connect \cref{eq:ZWO,eq:ZWO_total,eq:resistanceRandcapacitancC} to the charging of an electrolyte-filled pore, $R_p, R_{r}$, and $C$ must be expressed in terms of electrolyte properties and the pore and reservoir geometry.
We follow the choice of most authors and consider a cylindrical pore of radius $\varrho_p$ \cite{barcia2002application,cericola2016impedance,gassa1990electrochemical,jurczakowski2004impedance,keiser1976abschatzung,eloot1995calculation,kotz2000principles,song1999electrochemical} and a cylindrical reservoir of radius $\varrho_r$, so that $A^c_p=\pi \varrho_p^2$, $A^c_r=\pi \varrho_r^2$, and $A^s_p=2\pi\varrho_p\ell_p$.
However, we stress that the TL circuit may just as well be applied to pores and reservoirs with noncircular cross-sections. 
In this article, we will use the Poisson-Nernst-Planck equations to model the response of dilute electrolytes to small applied potentials. 
At steady state, this model yields the capacitance $c_{\rm EDL}=\varepsilon /\lambda_D$, where $\varepsilon$ is the electrolyte permittivity and where the Debye length $\lambda_D$ is the characteristic width of the equilibrium EDL.
Moreover, the PNP equations apply to electrolytes with a conductivity $\kappa=\varepsilon D/\lambda_D^2$, with $D$ being the ionic diffusivity, assumed to be equal among cations and anions.
Inserting all these expressions into \cref{eq:ZWO,eq:ZWO_total} seemingly gives us a theoretical impedance for arbitrary $\ell_p, \ell_r, \varrho_p, \varrho_r, D$, and $\lambda_D$.
This is not the case. 
As we explain below, underlying the derivation \cref{eq:ZWO,eq:ZWO_total} are several assumptions on the relation between these parameters, for instance, that the pore has a large aspect ratio ($\ell_p\gg\varrho_p$) and thin EDLs ($\varrho_p\gg\lambda_D$).

\subsection{Circuit models for porous electrode charging}
Several papers extended the TL model to account for, for instance, Faradaic processes \cite{levie1967electrochemical}, contact resistances, electrodes with resistance \cite{paasch1993theory}, and various pore shapes \cite{delevie1965,keiser1976abschatzung}.
$Z_p$ and the impedances of other TL-like circuits were also connected in ``super'' circuits to describe the charging of porous electrodes containing different-sized \cite{song1999electrochemical} or
hierarchically connected pores \cite{eikerling2005optimized,itagaki2010complex,gommes2022electrical}.
Others represented porous electrodes by a parallel connection of $m$ identical pores, for which the total impedance reads $Z=R_{r}+Z_p/m$ \cite{barcia2002application,orazem2008electrochemical,lasia2014electrochemical,cericola2016impedance}. 
Identifying $R_p/m=R_\text{tot}$ and $C m=C_\text{tot}$, however, yields 
\begin{equation}\label{eq:nparallelWarburg}
    Z=R_{r}+\sqrt{\frac{R_\text{tot}}{\im\omega C_\text{tot}} } \coth \sqrt{ \im\omega R_\text{tot}C_\text{tot} },
\end{equation}
that is, of the same form as \cref{eq:ZWO_total}, but with a different interpretation of its variables.

\Cref{eq:ZWO_total,eq:nparallelWarburg} having the same functional form signals a general problem of interpreting EIS data by equivalent circuits: fit parameters do not always have clear interpretations.
EIS on porous electrodes often yields data with shapes similar to the one in \cref{fig_Warburg} \cite{gassa1990electrochemical,jurczakowski2004impedance,ogihara2012theoretical,ogihara2015impedance,Li_2007,lust2004influence,lust2004influenceII}.
One can fit \cref{eq:nparallelWarburg} to such data, for instance, with impedance.py \cite{Murbach2020} or commercial software, or one can quickly estimate $R_\text{tot}\approx 3\{\Re[Z(\omega\to0)]-\Re[Z(\omega\to\infty)]\}$, $R_{r}\approx\Re[Z(\omega\to\infty)]$, and $C_\text{tot}\approx \pi^2/(2R_\text{tot}\omega^\star)$ from the limits and the 45-to-90-degrees transition of the complex plane plot.
Either way, while the complex plane plot of an electrode with thousands of pores may \emph{look} like that of \cref{eq:nparallelWarburg}, unless one has verified that all assumptions underlying its derivation are satisfied, it is unclear how the fit parameters $R_\text{tot}, R_{r}$, and $C_\text{tot}$ relate to the microscopic details of the system at hand.
By some independent experiment(s), one should thus determine the number of pores and their size and shape, verify that all pores are the same, verify that there are no hierarchical connections, etc.
Until that time, the fit parameter $R_\text{tot}$, for instance, is little more than a shorthand for $3\{\Re[Z(\omega\to0)]-\Re[Z(\omega\to\infty)]\}$
\footnote{If an electrode is indeed a bundle of $m$ parallel pores of known surface area $A^s_p$ per pore, one can estimate the number of pores $m=c_{\rm EDL}A^s_p/C_{\rm tot}$ using 
typical values for the specific capacitance $c_{\rm EDL}$.}. 

Another related problem of interpreting EIS spectra by equivalent circuits is that two circuits accounting for different mechanisms may have the same impedance.
Concretely, say one studies the effect of pore shape on porous electrode charging and that a particular complex plane plot can be fitted well by the equivalent circuit model of Keiser, Beccu, and Gutjahr \cite{keiser1976abschatzung} for the impedance of different shaped pores. 
Such a good fit, however, does not preclude some other straight-pore model, accounting for additional physical mechanisms, from fitting the same data.

\subsection{Microscopic models for single-pore charging}\label{sec:microscopicmodels}
While there is a historical tradition of interpreting EIS data through equivalent circuits, the above two examples showed some of their limitations.
Today's computational methods and resources allow one to predict EIS data through continuum models and molecular simulation \cite{tomlin2022impedance,mei2018physical,babel2018impedance,pireddu2023frequency}, which can capture hitherto neglected phenomena like image charge interaction, finite ion size, and nontrivial electrode geometries. 
The behavior of such complex systems might sometimes still be caught by equivalent circuits.
Still, it is better to start from a first-principles model and \emph{derive} its reduced-order behavior than to \emph{pose} an equivalent circuit model and view its fitting to data as a justification of the model itself.

Before one can understand the EIS response of electrodes containing thousands of intricately-connected different pores, one should understand the EIS response of model geometries.
In this regard, the mentioned single-pore PNP modeling studies \cite{sakaguchi2007,mirzadeh2014enhanced,henrique2021charging,henrique2022impact,yang2022simulating,tomlin2022impedance, alizadeh2017multiscale, henrique2021charging,aslyamov2022analytical} 
helped to verify and extend the classical circuit models of de Levie and his contemporaries.
In one of these works, we analytically solved the PNP equations for the charging of a single slit pore in contact with an electrolyte reservoir of negligible resistance \cite{aslyamov2022analytical}.
The case of small applied potentials and thin EDLs yielded an expression of the same form as the TL model's potential relaxation [viz. \cref{eq:poseyandmorozumi}]\footnote{Reference~\cite{henrique2021charging} analytically solved the PNP equations for the charging of a cylindrical pore. For thin EDLs, their solution also simplifies to TL model results}.
In place of the TL circuit's $R_p C $ appeared $\ell_p^2\lambda_D/(h_p D)$, with $h_p$ the pore's width.
The same expression results from multiplying the pore's capacitance $\mathcal{C}=\varepsilon h_p \ell_p/\lambda_D$ and electrolyte resistance $\mathcal{R}=\lambda_D^2 \ell_p/(\varepsilon Dh_p)$, both per unit length in the in-plane direction. 
Hence, in this case, there is an exact analytical correspondence between the microscopic 3d continuum model (PNP) and the reduced-order TL model, with an exact expression of the circuit parameters $R_p C $ in terms of electrode and electrolyte properties.
That means that, in this case, the fit parameters of the TL model relate unambiguously to microscopic electrode and electrolyte properties. 
Other PNP modeling studies focused on the step response of pores in contact with an electrolyte reservoir \cite{yang2022simulating,henrique2021charging,henrique2022impact}.
In these studies, the TL model predictions and the continuum data agreed decently but not precisely.

Despite these recent efforts, sixty years after de Levie's seminal papers, the charging of a single pore has still not been fully characterized.
Consider again the cylindrical electrolyte-filled pore of length $\ell_p$ and radius $\varrho_p$ filled with an electrolyte with a Debye length $\lambda_D$ and equal ionic diffusivities $D$, subject to a small sinusoidal voltage of angular frequency $\omega$ (ignore the electrolyte reservoir for now).
Of this model's four length scales, $\ell_p, \varrho_p, \lambda_D$, and $\sqrt{D/\omega}$, 12 dimensionless ratios can be constructed (more will enter when an electrolyte reservoir, finite ion size, etc. are introduced).
However, only three dimensionless ratios are independent; for instance, the Peclet-like parameter $\sqrt{D/\omega}/\ell_p$, the EDL overlap $\lambda_D/\varrho_p$, and the pore aspect ratio $\ell_p/ \varrho_p$. 
De Levie \cite{levie1963} implicitly discussed the product of the first two of these three ratios.
For small $\omega$, ions can keep up with the applied voltage, and EDLs are in quasi-equilibrium.
For large $\omega$, only the region near the pore mouth is charged and discharged.
Accordingly, when we solve for the time-dependent potential in the pore [viz. \cref{eq:psihatsolution}], we find that it varies over a frequency-dependent length $\ell_\omega=\ell_p/\sqrt{\im \omega R_p C }$ called the penetration depth \cite{levie1963}; hence, the dimensionless ratio $\ell_\omega/\ell_p$ determines the extent to which the pore is charged.
With \cref{eq:resistanceRandcapacitancC} and the expressions in the lines below it, we find $R_p C =2\ell_p^2\lambda_D/(\varrho_p D)$ and
\begin{equation}
    \frac{\ell_\omega}{\ell_p}=\sqrt{\frac{1}{2\im}\frac{D}{ \omega \ell_p^2}\frac{\varrho_p}{\lambda_D}}.
\end{equation}
Hence, $\ell_\omega/\ell_p$ is a product of two of the three mentioned dimensionless ratios.
The EDL overlap parameter was thus already implicit in de Levie's work.
Still, his results can only hold for $\varrho_p\gg\lambda_D$, as overlapping EDL correspond to finite in-pore potential values at late times, which cannot be captured by the TL circuit. 
EDL overlap has only recently been thoroughly addressed by Henrique, Zuk, and Gupta through analytical and numerical PNP calculations \cite{henrique2021charging,henrique2022impact}.
The third independent dimensionless ratio, the pore aspect ratio $\ell_p/ \varrho_p$, has been virtually unexplored \footnote{Reference~\cite{eloot1995calculationII} is a notable exception.}---so far, most equivalent circuit and PNP studies of pore charging (implicitly or explicitly) took $\ell_p/ \varrho_p\gg1$, for the following reason.
De Levie argued that, for the TL circuit to describe pore charging, equipotential lines in the electrolyte should be straight \cite{levie1963} and that short pores do not satisfy this condition (see page 372 of de Levie~\cite{levie1967electrochemical}).
The second box in \cref{fig0} shows equipotential lines based on numerical simulations described below (viz. \cref{sec:numericalstudy}).
This figure shows that equipotential lines are not straight near a finite-length pore's entrance.
This region will play a relatively larger role in the charging of short pores, so, indeed, the impedance of such pores cannot follow TL model predictions.

\subsection{Overview}
We comprehensively discuss single mesopore charging through ladder circuits and delineate by PNP modeling the validity of such circuits.
\Cref{sec:Fraction} shows that the pore impedance $Z_p$ can be analytically derived from its corresponding TL circuit---we also discuss several popular TL-circuit extensions. 
Our derivations entirely bypass the TL-type modeling usually employed.
\Cref{sec:circuit_to_ODE} reviews two ways to go from the different ladder circuits to their corresponding TL equations. 
In particular, we generalize Janssen~\cite{janssen2021transmission} to a case with Faradaic processes at the electrode surface.
In \cref{sec:ZfromTLEq}, we derive the impedances of different pore-reservoir systems from their corresponding TL equations using Posey and Morozumi's pore-reservoir boundary condition. 
In \cref{sec:stepresponse}, we relate a pore's impedance to its response to a step potential.
\Cref{sec:numericalstudy} presents numerical results for the PNP equations in blocking mesopores.
We determine the impedance of pore-reservoir systems with pores of different lengths and compare them to \cref{eq:ZWO_total}.
While we focus on single pore charging, we also discuss porous electrodes in \cref{sec:discussion}.
We conclude in \cref{sec:conclusion}.
In \cref{fig0}, we indicate in green the locations of the new contribution of this work.
We refer readers interested in practical applications of the TL model and its extensions to recent review papers \cite{huang2018diffusion,huang2020editors,moskon2021transmission} and textbooks \cite{conway2013electrochemical,orazem2008electrochemical,lvovich2012impedance,lasia2014electrochemical}.

\section{Impedance from circuits}\label{sec:Fraction}
\subsection{Standard TL circuit}\label{sec:reflectingwarburg}
\begin{figure}
    \includegraphics[width=\linewidth]{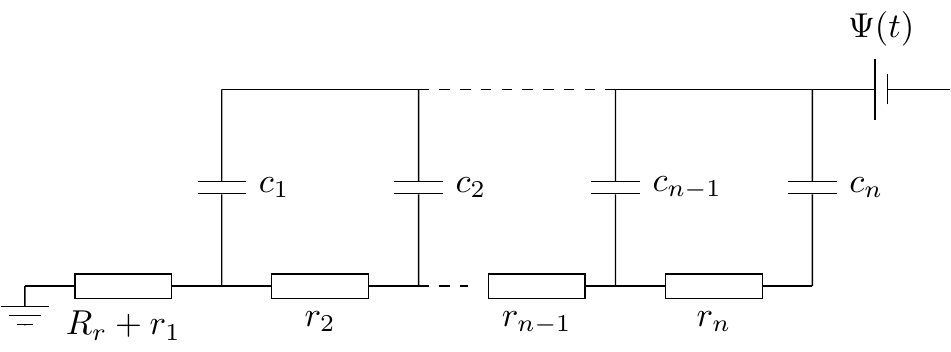}
    \caption{Standard TL circuit. \label{TLcircuit_standard}}
\end{figure}

The TL model partitions the resistance $R_p$ and capacitance $C$ of a pore into $n$ pieces of resistance $r_k$ and capacitance $c_k$, with $k=1,\ldots,n$.
These elements are then connected as shown in \cref{TLcircuit_standard}. 
The top line in this circuit represents the pore's metallic surface, which is subjected to a small sinusoidal potential $\Psi(t)=\Psi_0 \sin(\omega t)$.
The bottom row represents the electrolyte in the pore and in a reservoir of resistance $R_{r}$.

To determine the impedance of the circuit, we start at the last branch ($n$) and work our way to the reservoir resistor.
The impedance of the last ladder rung reads
\begin{equation}\label{eq:Z1}
    Z_n=r_n+\frac{1}{\im\omega c_n}\,.
\end{equation}
Likewise, the impedance of the $k$-th rung reads 
\begin{equation}\label{eq:Zi}
    Z_{k}=r_{k}+\frac{1}{\im\omega c_{k}+Z_{k+1}^{-1}},\qquad k =1,\ldots, n-1 \,.
\end{equation}
The impedance of the complete circuit is then $Z=R_{r}+Z_1$; note that $Z_1$ accounts for all ladder rungs.

The first-order rational difference equation \eqref{eq:Zi} previously appeared in Keiser, Beccu, and Gutjahr~\cite{keiser1976abschatzung}. 
That article considered noncylindrical pores, such that $r_k$ and $c_k$ varied along the circuit.
We consider here the simpler case of a straight and homogeneous pore, for which $c_{k}=c$ and $r_{k}=r$ and thus $R_p=rn$ and $C=cn$.
We rewrite \cref{eq:Z1,eq:Zi} with the scaled angular frequency $\bar{\omega} = \omega r c $ (throughout, bars indicate dimensionless quantities) and $Z_{k}=r a_{k}/b_{k}$, with $a_{k}$ and $b_{k}$ to be determined, to
\begin{align}\label{eq:pqratios}
    \frac{a_{n}}{b_{n}}&=\frac{\im \bar{\omega}+1}{\im \bar{\omega}}&\quad, \quad
    \frac{a_{k}}{b_{k}}&=\frac{(\im \bar{\omega}+1) a_{k+1} + b_{k+1}}{\im \bar{\omega} a_{k+1} + b_{k+1}}\,.
\end{align}
The same expressions result if one takes the ratio of the top and bottom elements of the following vectors,
\begin{subequations}\label{eq:matrixeq}
\begin{align}
    \begin{bmatrix}a_{n}\\b_{n}\end{bmatrix}&\propto B
    \begin{bmatrix}1\\ 0\end{bmatrix}\quad, \quad
    \begin{bmatrix}a_{k}\\b_{k}\end{bmatrix} \propto
    B
    \begin{bmatrix}a_{k+1}\\b_{k+1}\end{bmatrix}\,,\\
    &\text{with}\qquad B\equiv\begin{bmatrix}\im \bar{\omega}+1 &1 \\ \im \bar{\omega} & 1\end{bmatrix}\,.\label{eq:matrixeqB}
    \end{align}
\end{subequations}
\Cref{eq:matrixeq} implies that 
\begin{equation}\label{eq:ptoq}
    \begin{bmatrix}a_{1}\\b_{1}\end{bmatrix} \propto B^{n} \begin{bmatrix}1\\ 0\end{bmatrix}\,.
\end{equation}
By diagonalizing $B$ as $Bu_\pm = \lambda_\pm u_\pm$, where
\begin{equation}
    \lambda_{\pm} = 1+\frac{\im \bar{\omega}}{2} \pm \sqrt{\im \bar{\omega}-\frac{\bar{\omega}^2}{4} },\quad
    u_{\pm} = \begin{bmatrix}\lambda_{\pm}-1\\\im \bar{\omega}\end{bmatrix}\,,
\end{equation}
and by using $B^{n} =PD^{n} P^{-1}$, where $P= \begin{bmatrix} u_+ &u_- \end{bmatrix}$, we rewrite \cref{eq:ptoq} to
\begin{align}
    \begin{bmatrix}a_{1}\\b_{1}\end{bmatrix}&\propto
    \begin{bmatrix}\lambda_{+}-1 &\lambda_{-} -1\\\im \bar{\omega} & \im \bar{\omega}\end{bmatrix}
    \begin{bmatrix}\lambda_{+}^{n} &0\\ 0& \lambda_{-}^{n}\end{bmatrix}
    \begin{bmatrix}\im \bar{\omega} &1-\lambda_{-}\\-\im \bar{\omega} & \lambda_{+}-1 \end{bmatrix}
    \begin{bmatrix}1\\ 0\end{bmatrix}\nn
    &=\begin{bmatrix}(\lambda_{+}-1)\lambda_+^n-(\lambda_--1)\lambda_-^n\\ \im \bar{\omega}(\lambda_{+}^{n}-\lambda_{-}^{n}) \end{bmatrix}.
\end{align}
Hence, $Z_1=r a_1/b_1$ amounts to 
\begin{equation}\label{eq:Zn}
    Z_1 =\frac{n}{\im \omega C}\frac{(\lambda_{+}-1)\lambda_+^n-(\lambda_--1)\lambda_-^n}{\lambda_{+}^n -\lambda_{-}^n}\,.
\end{equation}
Next, using $\bar{\omega} = \omega R_p C /n^2$, we rewrite the eigenvalues to
\begin{equation}
    \lambda_{\pm} = 1 \pm \frac{\sqrt{\im\omega R_p C }}{n }+O\left(\frac{1}{n^2}\right)\,,
\end{equation}
which, inserted into \cref{eq:Zn}, yields
\begin{equation}
    Z_1 =\sqrt{\frac{R_p }{\im\omega C} }\frac{\lambda_{+}^n + \lambda_{-}^n}{\lambda_{+}^n - \lambda_{-}^n}+O\left(\frac{R_p}{n^2}\right)\,.
\end{equation}
Using that $\lim_{n\to\infty} (1\pm x/n)^n = \exp\left( \pm x \right)$, we find
\begin{equation}
    \lim_{n\to\infty} Z_1 =\sqrt{\frac{R_p }{\im\omega C} } \coth \sqrt{ \im\omega R_p C },
\end{equation}
that is, $Z_p$ [\cref{eq:ZWO}].
The impedance of the circuit, including the reservoir resistance, then amounts to \cref{eq:ZWO_total}.

\subsection{Contact resistance}
\begin{figure}
    \includegraphics[width=\linewidth]{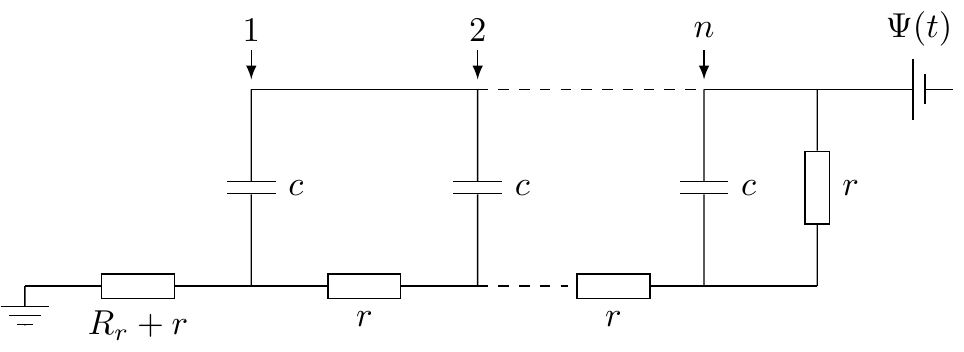}
    \caption{TL circuit with contact resistance. \label{TLcircuit_contact_resistance}}
\end{figure}

To account for the resistance between a porous electrode and a current collector, we extend the TL circuit with a resistor of resistance $r$ in the ladder's last rung, see \cref{TLcircuit_contact_resistance}.
The impedance of the last rung now reads 
\begin{equation}\label{eq:Z1_transmissive}
    Z_n=r+\frac{1}{\im\omega c+r^{-1}}\,.
\end{equation}
We rewrite \cref{eq:Z1_transmissive} to 
\begin{align}
    &\frac{a_{n}}{b_{n}} =\frac{\im \bar{\omega}+2}{\im \bar{\omega}+1}
    &\quad\iff\quad
    \begin{bmatrix}a_{n}\\b_{n}\end{bmatrix}\propto B
    \begin{bmatrix}1\\ 1\end{bmatrix}\,,
\end{align}
with the same $B$ as in \cref{eq:matrixeqB}.
Instead of \cref{eq:ptoq}, now 
\begin{equation}
    \begin{bmatrix}a_{1}\\b_{1}\end{bmatrix}\propto B^{n} \begin{bmatrix}1\\ 1\end{bmatrix}\,,
\end{equation}
which yields
\begin{align}
    &Z_1 = \frac{n}{\im \omega C}\times\\
    &\frac{\lambda_+^n(\lambda_+-1)(1-\lambda_-+\im \bar{\omega})+\lambda_-^n(\lambda_--1)(\lambda_+-1-\im \bar{\omega})}{\lambda_{+}^{n}(1-\lambda_-)+\lambda_{-}^{n}(\lambda_+-1)}\nonumber\,,
\end{align}
and, in turn, 
\begin{equation}\label{eq:warburgshort}
    \lim_{n\to\infty} Z_1 =\sqrt{\frac{R_p }{\im\omega C} } \tanh \sqrt{ \im\omega R_p C  }\,,
\end{equation}
which we denote $Z_{\rm con}$ from hereon.
The total resistance thus reads
\begin{equation}\label{eq:ZWS_total}
    \frac{Z(\omega)}{R_p}=\frac{R_{r}}{R_p}+\frac{\tanh\sqrt{\im \omega R_p C }}{\sqrt{\im \omega R_p C }}\,.
\end{equation}

\subsection{Faradaic processes at electrode-electrolyte interface}
\begin{figure}
    \includegraphics[width=\linewidth]{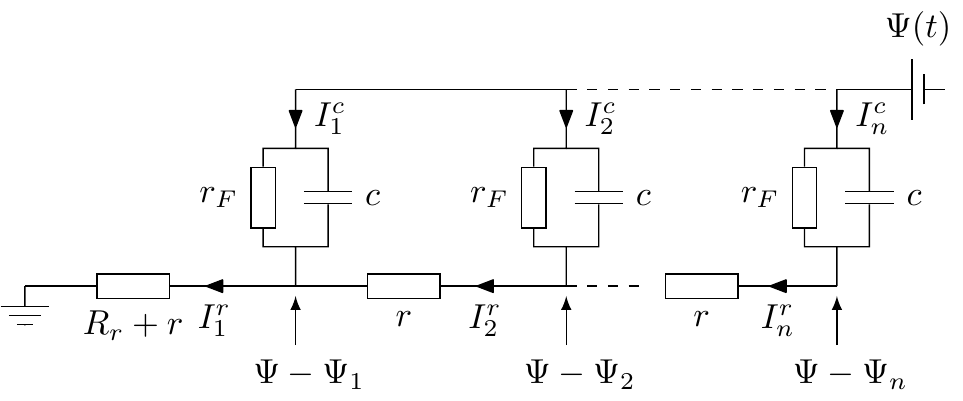}
    \caption{``Leaky'' TL circuit for a pore with both capacitive and Faradaic charging. \label{TLcircuit_surface_conduction}}
\end{figure}

The circuit in \cref{TLcircuit_surface_conduction} models a pore with Faradaic (charge transfer) currents at its surface \cite{levie1967electrochemical, 2007Itagaki,ogihara2012theoretical,ogihara2015impedance} and no dc gradients in potential and ion concentrations \cite{keddam1984impedance,lasia1995impedance}. 
The associated charge transfer resistance $R_F$ is partitioned into $n$ pieces so that $R_F=r_F/n$.
(The same circuit is used in the EIS analysis of solar cells and thin film diffusion \cite{bisquert2002}; in that context, $R_F$ is the recombination resistance.)
In this case,
\begin{subequations}\label{eq:faradaic}
\begin{align}
    Z_n&=r+\frac{1}{\im\omega c+r_F^{-1}}\,,\label{eq:Z1_faradaic}\\
    Z_{k}&=r+\frac{1}{\im\omega c+r_F^{-1}+Z_{k+1}^{-1}},\qquad k=1,\dots,n-1\,.\label{eq:Zi_faradaic}
\end{align}
\end{subequations}
With $\gamma=r/r_F$, \cref{eq:faradaic} reduces to
\begin{subequations}
\begin{align}
    \frac{a_{n}}{b_{n}}&= \frac{\im \bar{\omega}+\gamma+1}{\im \bar{\omega}+\gamma }\,,\\
    \frac{a_{k}}{b_{k}}&= \frac{(\im \bar{\omega}+\gamma+1) a_{k+1} + b_{k+1}}{(\im \bar{\omega}+\gamma) a_{k+1} + b_{k+1}}\,,
\end{align}
\end{subequations}
By writing $\im \bar{\omega}'=\im \bar{\omega}+\gamma$ and dropping primes, we recover \cref{eq:pqratios}.
Hence, \cref{eq:Zn} again holds, but the eigenvalues are now 
\begin{equation}
    \lambda_{\pm}= 1 \pm \frac{ \sqrt{\im\omega R_p C +R_p/R_F}}{n }+O\left(\frac{1}{n^2}\right),
\end{equation}
where we used that $\gamma=r/r_F=(R_p/n) /(R_F n) =O(n^{-2})$.
We thus find
\begin{equation}
    \lim_{n\to\infty} Z_1 =\sqrt{\frac{R_p R_F }{1+\im\omega R_F C} } \coth \sqrt{\frac{R_p}{R_F}(1+ \im\omega R_F C) }\,,
\end{equation}
which is implicit in Eqs. (96), (103), and (104) of de Levie~\cite{levie1967electrochemical} and which we call the Faradaic pore impedance $Z_{F}$ from hereon.

The total resistance thus reads
\begin{equation}\label{eq:ZWfaradaic_total}
    \frac{Z(\omega)}{R_p}=\frac{R_{r}}{R_p}+\frac{\coth \sqrt{\frac{R_p}{R_F}+ \im\omega R_p C }}{\sqrt{\frac{R_p}{R_F}+ \im\omega R_p C }}\,.
\end{equation}

\begin{figure}
    \includegraphics[width=\linewidth]{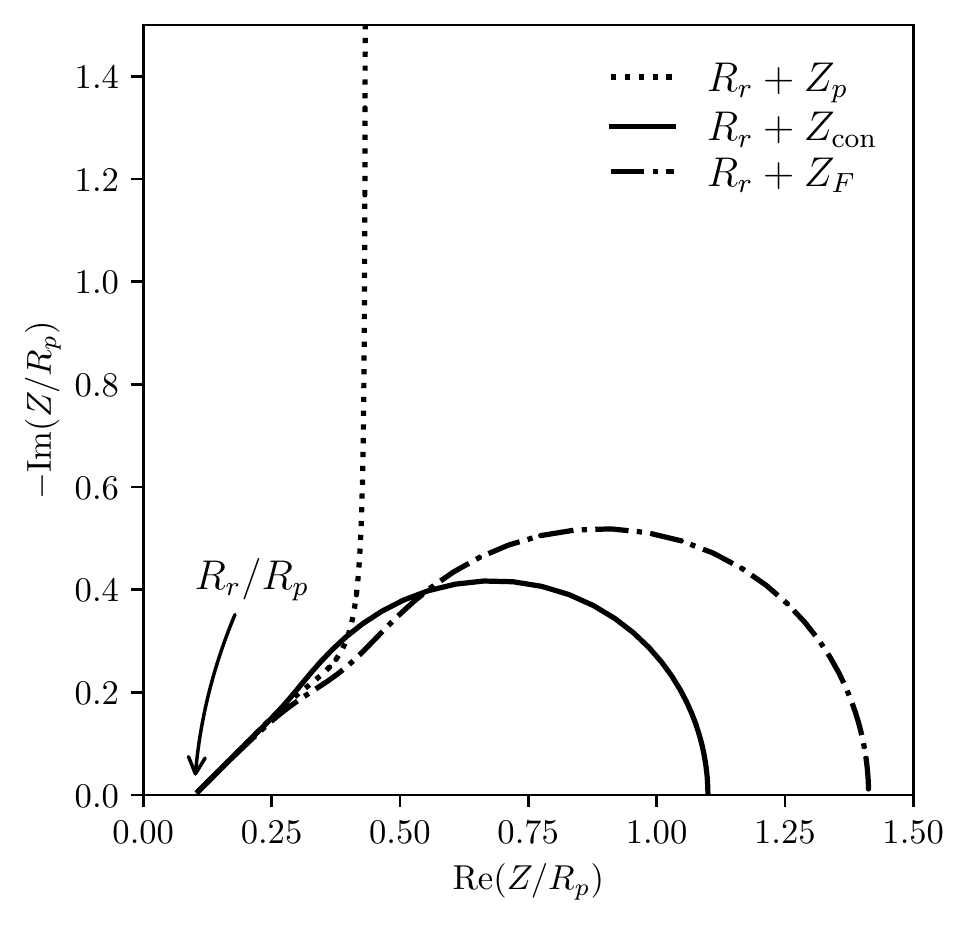}
    \caption{
    Complex plane plot of a reservoir resistor connected to the pore impedance $Z_p$ [\cref{eq:ZWO_total}, dotted] and extensions of the TL circuit accounting for a contact resistance, $Z_{\rm con}$ [\cref{eq:ZWS_total}, full line] and Faradaic processes, $Z_{F}$ [\cref{eq:ZWfaradaic_total}, dash-dotted].
    We set $R_p/R_{r}=10$ and $R_p/R_F=1$. 
    \label{fig_collection}}
\end{figure}

\Cref{fig_collection} shows a complex plane plot of \cref{eq:ZWO_total,eq:ZWS_total,eq:ZWfaradaic_total} for $R_p/R_{r}=10$ and $R_p/R_F=1$.

\subsection{Further extensions}
\subsubsection{Ladders with large rungs}
In his famous lectures, Feynman derived the impedance of an infinite $LC$ ladder \cite{feynman2011feynman}. 
Feynman argued that, for large $k$, the impedance of successive rungs should be the same: $Z_{k+1}=Z_k$.
Barbero and Lelidis repeated this analysis for the TL circuit \cite{barbero2017analysis} with infinitely many $R$ and $C$ elements.
Replacing $r\to R$ and $c\to C$ and setting $Z_{k+1}=Z_k$ in \cref{eq:Zi} yields $Z_k^2-RZ_k-R/(\im \omega C)=0$. 
The positive root of this quadratic equation reads $Z_k=R/2+R\sqrt{1/4+1/(\im \omega R C )}$ \footnote{\cit{barbero2017analysis} seems to have a minus sign error in their corresponding Eq. (45)}, which, for small $\omega R C $, displays Warburg-like scaling $Z\propto R/2+\sqrt{R/(\im \omega C)}$.
The crucial difference between the analyses of Barbero and Lelidis \cite{barbero2017analysis} and our derivation in \cref{sec:reflectingwarburg} is that we consider a pore whose overall resistance $R_p$ and capacitance $C$ are fixed (and finite)---taking $n\to\infty$, the resistors $r=R_p/n$ and capacitors $c=C/n$ in our circuit become ever smaller. 
We can recover Barbero's result by replacing all $r\to R$ and $c\to C$.
In that case, \cref{eq:Zn} changes to
\begin{equation}
    Z_1 =\frac{1}{\im \omega C}\frac{(\lambda_{+}-1)\lambda_+^n-(\lambda_--1)\lambda_-^n}{\lambda_{+}^n - \lambda_{-}^n}\,,
\end{equation}
with 
\begin{equation}
    \lambda_{\pm} = 1+\frac{\im \omega R C }{2} \pm \sqrt{\im \omega R C -\frac{(\omega R C)^2}{4} }.\label{eq:lambda_Barbero}
\end{equation}
We write these eigenvalues in polar form, $\lambda_{\pm}=|\lambda_\pm|\me^{\im \varphi_\pm}$ with $\varphi_\pm$ being the arguments of the complex $\lambda_{\pm}$; hence, $\lambda_\pm^n=|\lambda_\pm|^n\me^{\im n \varphi_\pm}$.
From \cref{eq:lambda_Barbero}, one finds
\begin{align}
    |\lambda_+|^2-|\lambda_-|^2&\geq\sqrt{\frac{4\omega R C +(\omega R C )^3}{2}}\big[16+(\omega R C)^2\big]^{1/4}\nn
    &\geq 0,\,
\end{align}
where the equality holds for $\omega=0$. 
Hence, $|\lambda_+|>|\lambda_-|$ for $\omega >0$, which implies that, for $\omega >0$, 
\begin{equation}
     \lim_{n\to\infty} Z_1 =\frac{1}{\im \omega C}(\lambda_{+}-1) = \frac{R}{2}+\frac{R}{2} \sqrt{1+\frac{4}{\im \omega R C }}\,,
\end{equation}
in agreement with the result obtained by Feynman's method.

\subsubsection{Distributed inductance}
The derivation in \cref{sec:reflectingwarburg} allows us to study an $LC$ network, not with finite $L$ and $C$ elements like Feynman did but with an overall $L$ and $C$ distributed over $n$ elements, such that, again, $C=cn$ and now also $L=ln$.
Replacing the small resistors of \cref{TLcircuit_standard} with small inductors of impedance $\im \omega l$, we can again use \cref{eq:ZWO} but replace $R_p\to \im \omega L$, hence,
$Z=\sqrt{L/C } \coth \sqrt{-\omega^2 LC}$, in agreement with Eq.~(66) of Barbero and Lelidis \cite{barbero2017analysis}.

\subsubsection{Electrode resistance}
\begin{figure}
    \includegraphics[width=\linewidth]{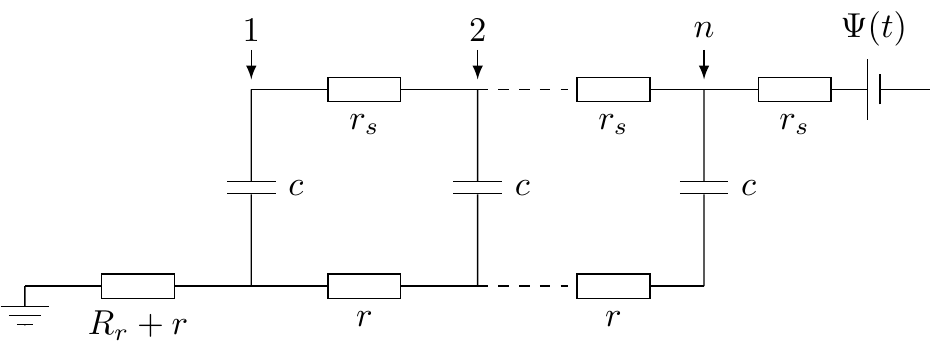}
    \caption{TL circuit with electrode resistance. \label{TLcircuit_surface_resistance}}
\end{figure}

Paasch, Micka, and Gersdorg \cite{paasch1993theory} studied a transmission line with resistances in both channels, see \cref{TLcircuit_surface_resistance}.
Such a circuit corresponds to a case where not only the electrolyte but also the electrode has a finite resistance, $R_s=n r_s$, with $r_s$ the small resistance of the elements in the circuit.
To derive a recursion relation like \cref{eq:Zi} for this circuit probably requires repeated use of Y-$\Delta$ transformations.
We have not yet been able to do so, so we leave this problem for future research.

\subsubsection{Pores with varying section}
Keiser, Beccu, and Gutjahr numerically solved the recursion relation \cref{eq:Zi} for pores with varying sections, for which $c_k$ and $r_k$ in \cref{TLcircuit_standard} are not constant along the circuit \cite{keiser1976abschatzung}.
Analytically solving \cref{eq:Zi} for pores with varying sections will be difficult. 
$Z_k$ can still be written as \cref{eq:matrixeq}, but $\im \bar{\omega}$ will depend on $k$. 
Hence, a product of $k$ different matrices will appear, and we can no longer use $B^k=PD^k P^{-1}$.
Progress may be possible for the particular case of a groove, for which de Levie found an analytical expression \cite{delevie1965}.

\section{From circuits to differential equations}\label{sec:circuit_to_ODE}
We review two ways of extracting a TL equation from its corresponding equivalent circuit. 
We focus on the leaky TL circuit (\cref{TLcircuit_surface_conduction}) for concreteness.

\begin{figure}
    \includegraphics[width=0.9\linewidth]{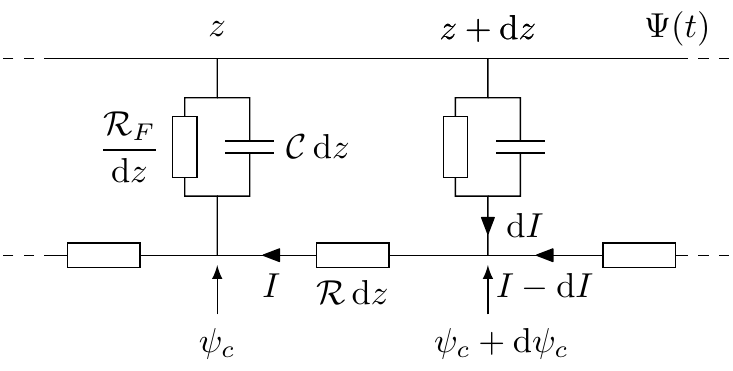}
    \caption{``Leaky'' TL circuit for a pore with capacitive and Faradaic charging, with capacitance and resistances per unit length. 
    \label{deLeviecircuit}}
\end{figure}

\subsection{De Levie's argument}
De Levie's derivation \cite{levie1963} of the TL equation goes as follows.
\Cref{deLeviecircuit} is a zoom-in of \cref{TLcircuit_surface_conduction} without a specified start or end. 
Again, the top line in this circuit represents the electrode, which is at $\Psi(t)$ everywhere.
The bottom row represents the electrolyte phase, which has a centerline potential $\psi_c$ that varies along the pore.
The voltage drop $\diff \psi_c$ over a differential resistor is 
\begin{align}\label{eq:dpsi}
    \diff \psi_{c}&=\frac{\partial \psi_{c}}{\partial z}\diff z=-I\mathcal{R} \diff z\nn
    \frac{\partial \psi_{c}}{\partial z}&=-I\mathcal{R}\,,
\end{align}
with $\mathcal{R}$ being the electrolyte resistance per unit length.
For $\psi_{c}$ increasing in the $z$ direction, the electric field and, hence, the  ionic current point in the $-z$ direction, explaining the minus sign in \cref{eq:dpsi}.

\begin{figure}
    \includegraphics[width=0.9\linewidth]{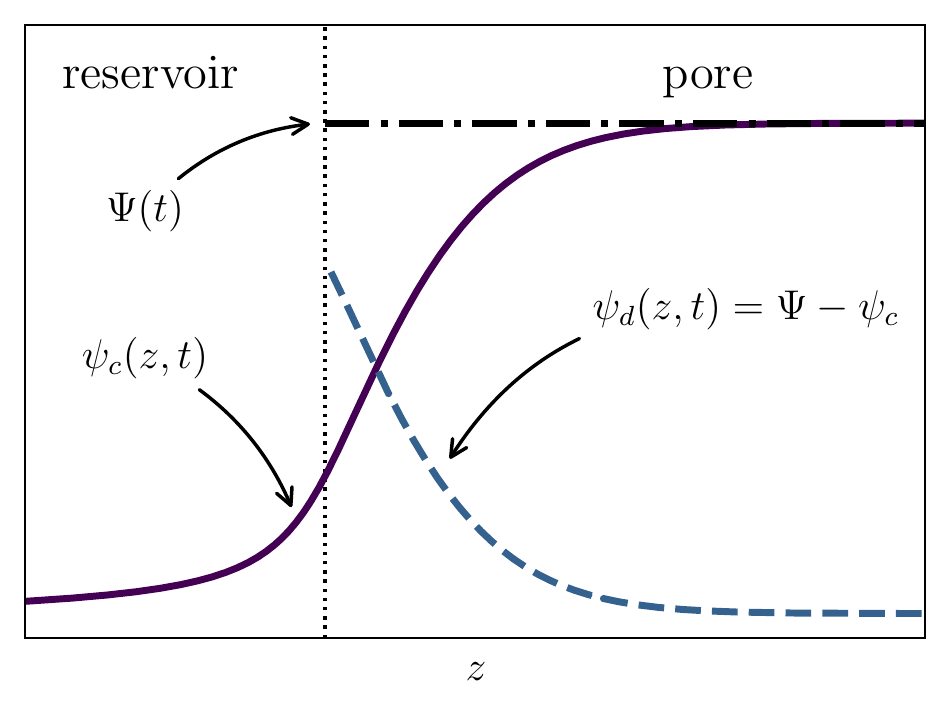}
    \caption{
    Schematic showing the centerline potential $\psi_c(z,t)$ and potential drop $\psi_d(z,t)$ in a reservoir-pore system. 
    The TL model only models the pore region and accounts for the reservoir through a boundary condition.
    The centerline potential is drawn here with numerical data from \cref{sec:numericalstudy}. 
    \label{fig:centerline}}
\end{figure}

For expressing the current $\diff I$ that flows into the bottom line in \cref{deLeviecircuit}, it is useful to introduce the potential drop $\psi_d(z,t)=\Psi(t)-\psi_c(z,t)$ between the pore wall, which is at $\Psi(t)$, and the center of the pore, which is at $\psi_c(z,t)$; see \cref{fig:centerline}.
The current that goes into a parallel-connected resistor and capacitor, with infinitesimal resistance  $\mathcal{R}_F/\diff z$ and capacitance $\mathcal{C}\diff z$, respectively, then reads
\begin{align}\label{eqdIdz}
    \diff I&=\mathcal{C} \diff z \frac{\partial \psi_{d}}{\partial t}+\frac{\psi_{d}}{\mathcal{R}_F} \diff z\nn
    \frac{\partial I}{\partial z}&=\mathcal{C} \frac{\partial \psi_{d}}{\partial t}+\frac{\psi_{d}}{\mathcal{R}_F}\,.
\end{align}
Rewriting \cref{eq:dpsi} in terms of $\psi_d$, taking a $z$ derivative, and inserting \cref{eqdIdz}, we find
\begin{equation}\label{eq:TLeq}
    \mathcal{RC} \frac{\partial \psi_{d}}{\partial t}=\frac{\partial^2 \psi_{d}}{\partial z^2}-\psi_{d}\frac{\mathcal{R}}{\mathcal{R}_F}\,,
\end{equation}
which is the TL equation for a case with homogeneous Faradaic surface conduction.

Once we introduce the pore's length $\ell_p$, we can express the per-unit-length resistances and capacitance, $\mathcal{R}=R_p/\ell_p$, $\mathcal{R}_{F}=R_F\ell_p$, and $\mathcal{C}=C/\ell_p$.
Still, the downside of the above argument is that, while it yields the correct TL equation, it does not inform on the boundary conditions that should be used.
As a result, different authors solved the TL equation for different boundary conditions.
Conversely, drawing a particular circuit including the first and last rungs of the ladder (like we did in \cref{TLcircuit_standard,TLcircuit_contact_resistance,TLcircuit_surface_conduction,TLcircuit_surface_resistance}) fixes the boundary conditions---as we will show below, there is no room for variations.

\subsection{Ref.~\cite{janssen2021transmission} argument}
One of us~\cite{janssen2021transmission} showed how the TL equation, including its boundary conditions [viz. \cref{eq:TLeqPosey}] can be directly related to the TL circuit.
The argument given there revolved around a finite-difference expression of the TL equation, including correct boundary conditions, which, in the limit $n\to\infty$ is identical to a matrix differential equation that can also be derived directly from the TL circuit.
Here, we repeat the argument for the slightly more involved circuit in \cref{TLcircuit_surface_conduction} (and also shortly discuss the case of a circuit with a contact resistance, see \cref{TLcircuit_contact_resistance}).

\subsubsection{Combining Ohm's and Kirchhoff's laws for all rungs of a ladder circuit}
For the circuit in \cref{TLcircuit_surface_conduction}, Ohm's law states that 
\begin{subequations}\label{eq:Ohmleaky}
\begin{align}
    I^{r}_{1}(t) (R_{r}+r) &=\Psi(t)-\Psi_{1}(t), \label{eq:Ohmleaky2}\\
    I^{r}_{k}(t) r &=\Psi_{k-1}(t)-\Psi_{k}(t), \qquad k=2,\ldots,n,\label{eq:Ohmleaky1}
\end{align}
\end{subequations}
with $\Psi(t)$ the potential of an external voltage source, $\Psi_{k}$ the potential drop over the $k$-th rung of the ladder, and $I^{r}_{k}(t)$ the current through the $k$-th resistor. 
Kirchhoff's junction rule gives
\begin{subequations}\label{eq:Kirchhoffleaky}
\begin{align}
    I^{c}_{k}(t)&=I^{r}_{k}(t)-I^{r}_{k+1}(t), \qquad k=1,\ldots,n-1,\label{eq:Kirchhoffleaky2}\\
    I^{c}_{n}(t)&=I^{r}_{n}(t)\,.\label{eq:Kirchhoffleaky1}
\end{align} 
\end{subequations}
Now, the current into the $k$-th rung reads 
\begin{equation}\label{eq:leakysurface}
    I^{c}_{k}(t)=\frac{1}{r_F}\Psi_{k}(t)+c\dot{\Psi}_{k}(t), \qquad k=1,\ldots,n,
\end{equation}
where $\dot{\Psi}_{k}(t)$ is the time derivative of the voltage drop across this rung. 
The above setup deviates from our previous work \cite{janssen2021transmission} in two places.
First, the $1/r_F$ term on the right-hand side in \cref{eq:leakysurface} was absent in \cit{janssen2021transmission} as we neglected surface conduction there.
Second, the circuit in \cit{janssen2021transmission} contained $R_{r}$ rather than $R_{r}+r$ in the leftmost resistor.
As a result, its Ohm's law corresponding to \cref{eq:Ohmleaky2} did not contain $r$.
For consistency with \cref{sec:Fraction}, we maintain this $r$.

Combining \cref{eq:leakysurface,eq:Kirchhoffleaky2} gives
\begin{align}
    \frac{1}{r_F}\Psi_{k}(t)+c\dot{\Psi}_{k}(t)&=I^{r}_{k}(t)-I^{r}_{k+1}(t), \nn
    &\qquad\qquad k=1,\ldots,n-1.
\end{align}
Next, inserting \cref{eq:Ohmleaky1} gives, for $k=2,\ldots,n-1$,
\begin{equation}\label{eq:psik}
    \frac{r}{r_F}\Psi_{k}(t)+rc\dot{\Psi}_{k}(t)=\Psi_{k-1}(t)-2\Psi_{k}(t)+\Psi_{k+1}(t).
\end{equation}
For $k=1$, we have to insert \cref{eq:Ohmleaky2} instead, giving,
\begin{align}\label{eq:psi1}
    \frac{r}{r_F}\Psi_{1}(t)+rc\dot{\Psi}_{1}(t)&=\frac{r}{R_{r}+r}\Psi(t)-\left(1+\frac{r}{R_{r}+r}\right)\Psi_{1}(t)\nn
    &\quad+\Psi_{2}(t)\,.
\end{align}
Finally, for $k=n$, we combine \cref{eq:leakysurface,eq:Kirchhoffleaky1,eq:Ohmleaky1} and find
\begin{equation}\label{eq:psin}
    \frac{r}{r_F}\Psi_{n}(t)+rc\dot{\Psi}_{n}(t)=\Psi_{n-1}(t)-\Psi_{n}(t)\,.
\end{equation}
By writing $\boldsymbol{\Psi}(t)= \left[\Psi_{1}(t),\ldots, \Psi_{n}(t)\right]^{\intercal}$ and $\mathbf{e}_{1}= \left[1,0,\ldots,0\right]^{\intercal}$, we can now collect \cref{eq:psik,eq:psi1,eq:psin} into the following matrix differential equation:
\begin{subequations}\label{eq:TLcircuiteq}
\begin{align}
    &R_p C\dot{\boldsymbol{\Psi}}(t)=\frac{n^2 R_p}{n R_{r}+R_p} \Psi(t) \mathbf{e}_{1}+ n^2 M_1 \boldsymbol{\Psi}(t)-\frac{R_p}{R_F } \boldsymbol{\Psi}(t)\,,\label{eq:matrixeq2}\\
    &M_1=
    \begin{bmatrix}
    -1-r/(R_{r}+r)& 1& &&\\
    1& -2& 1&& \\
    &\ddots & \ddots &\ddots &\\
    & & 1& -2&1\\
    & & & 1&-1\\
    \end{bmatrix}\,.\label{eq:matrixM1}
\end{align}
\end{subequations}
The matrix $M_1\in \mathbb{R}^{n\times n}$ can be diagonalized analytically, with its eigenvalues and eigenvectors expressed using Chebyshev polynomials.
\Cref{eq:TLcircuiteq} can thus be solved analytically, with its solution expressed in terms of these eigenvalues and eigenvectors \cite{janssen2021transmission}.

\subsubsection{Finite-difference formulation of TL equation}
In the limit $n\to\infty$, \cref{eq:TLcircuiteq} turns out to be equal to a finite difference scheme of the following equation: 
\begin{subequations}\label{eq:TLeqfaradaic}
\begin{align}
    R_p C  \partial_{t}\psi_d&=\ell_p^2 \partial_{z}^{2}\psi_d-\frac{R_p}{R_F}\psi_d\,, & &z\in[0,\ell_p]\,,\\
    \psi_d(z,0)&=0\,, \label{eq:TLinit}\\
    \ell_p\partial_{z}\psi_d(0,t)&=\frac{R_p}{R_{r}}[\psi_d(0,t)-\Psi(t)], \\
    \partial_{z}\psi_d(\ell_p,t)&=0\,.\label{eq:TLbcOut}
\end{align}
\end{subequations}
To show the connection between \cref{eq:TLeqfaradaic,eq:TLcircuiteq}, we discretize $z$ but not $t$.
Partitioning $[0,\ell_p]$ into $m$ pieces of width $h = \ell_p/m$ yields a uniform grid of $m+1$ gridpoints, at $z_{i}=ih$ with $i\in\{0,\ldots,m\}$.
On these gridpoints, the continuous electrostatic potential is approximately $\psi_{i}=\psi_d(z_{i})$.
A central difference approximation now gives $ \partial_{z}^{2}\psi_d(z_{i})\simeq (\psi_{i-1}-2\psi_{i}+\psi_{i+1})/h^2$.
To implement the Robin boundary condition at $z=0$, we introduce a ghost grid point at $z=-h$ and corresponding $\psi_{-1}$.
Now, approximating the $z$-derivative through a backward difference $\partial_{z}\psi(0)\simeq (\psi_{0}-\psi_{-1})/h$, the Robin boundary condition yields $\psi_{-1}=\psi_{0}+\xi[\Psi(t)-\psi_{0}]/m$, with $\xi=R_p/R_{r}$.
Similar reasoning and a forward difference yield $\psi_{m+1}=\psi_{m}$ for the Neumann condition \cite{strang2014functions}.
After grouping the above expressions and writing $\boldsymbol{\psi}(t)= \left[\psi_{0}(t),\ldots, \psi_{m}(t)\right]^{\intercal}$, \cref{eq:TLeqfaradaic} is approximated by
\begin{subequations}\label{eq:findiff}
\begin{align}
    R_p C  \dot{\boldsymbol{\psi}}(t)&=m\frac{R_p}{R_{r}} \Psi(t) \mathbf{e}_{1}+ m^2 M_2 \boldsymbol{\psi}(t)- \frac{R_p}{R_F } \boldsymbol{\psi}(t)\,,\\
    M_2&=
    \begin{bmatrix}
    -1-r/R_{r}& 1& &&\\
    1& -2& 1&& \\
    &\ddots & \ddots &\ddots &\\
    & & 1& -2&1\\
    & & & 1&-1\\
    \end{bmatrix}\,,\label{eq:matrixM2}
\end{align}
\end{subequations}
with $M_2\in \mathbb{R}^{m+1\times m+1}$.
After setting $m+1=n$, differences between \cref{eq:findiff,eq:TLcircuiteq} are of subleading order in $n$. 
In \cit{janssen2021transmission}, where we did not add $r$ to the reservoir resistance [\cref{eq:Ohmleaky1}], we had $M_1=M_2$.
Still, differences subleading in $n$ between the prefactors on the right-hand sides of \cref{eq:findiff,eq:TLcircuiteq} remained for that choice as well.

\subsubsection{TL equation for TL circuit with contact resistance}
We can use the above arguments to find the corresponding equations for the circuit with a contact resistance, \cref{TLcircuit_contact_resistance}.
In this case, we should omit the $\Psi_k(t)/r_F$ term from \cref{eq:psik} and set $r_F=r$ in \cref{eq:psin}.
We find 
\begin{subequations}\label{eq:TLshortmatrixeq}
\begin{align}
    &R_p C\dot{\boldsymbol{\Psi}}(t)=\frac{n^2 R_p}{n R_{r}+R_p}  \Psi(t) \mathbf{e}_{1}+ n^2 M_3 \boldsymbol{\Psi}(t)\,,\label{eq:matrixeq4}\\
    &M_3=
    \begin{bmatrix}
    -1-r/(R_{r}+r)& 1& &&\\
    1& -2& 1&& \\
    &\ddots & \ddots &\ddots &\\
    & & 1& -2&1\\
    & & & 1&-2\\
    \end{bmatrix}\,,\label{eq:matrixM3}
\end{align}
\end{subequations}
with $M_3\in \mathbb{R}^{n\times n}$.
Similar to the above, we can show that \cref{eq:matrixM3} corresponds to 
\begin{subequations}\label{eq:TLeqshort}
\begin{align}
    R_p C  \partial_{t}\psi_d&=\ell_p^2 \partial_{z}^{2}\psi_d\,, & &z\in[0,\ell_p]\,,\\
    \psi_d(z,0)&=0\,, \label{eq:TLshortinit}\\
    \ell_p\partial_{z}\psi_d(0,t)&=\frac{R_p}{R_{r}}[\psi_d(0,t)-\Psi(t)], \label{eq:TLbcin}\\
    \psi_d(\ell_p,t)&=0\,.\label{eq:TLbcshortOut}
\end{align}
\end{subequations}
Using the same notation as before, in a finite difference scheme of \cref{eq:TLeqfaradaic}, the boundary conditions \cref{eq:TLbcin,eq:TLbcshortOut} reduce to $\psi_{-1}=\psi_{0}+\xi[\Psi(t)-\psi_{0}]/m$ and $\psi_{m}=0$, respectively. 
The latter condition modifies the finite difference for the second derivative at $z_{m-1}$ as $\partial_z^2\psi_d(z_{m-1})\approx \psi_{m-2}-2\psi_{m-1}$.
Combining the non-zero values $\psi_i$ in the vector $\boldsymbol{\psi}(t)= \left[\psi_{0}(t),\ldots, \psi_{m-1}(t)\right]^{\intercal}$ one can approximate \cref{eq:TLeqshort} as
\begin{subequations}\label{eq:findiff-2}
\begin{align}
    R_p C  \dot{\boldsymbol{\psi}}(t)&=m\frac{R_p}{R_{r}} \Psi(t) \mathbf{e}_{1}+ m^2 M_4 \boldsymbol{\psi}(t)\,,\\
    M_4&=
    \begin{bmatrix}
    -1-r/R_{r}& 1& &&\\
    1& -2& 1&& \\
    &\ddots & \ddots &\ddots &\\
    & & 1& -2&1\\
    & & & 1&-2\\
    \end{bmatrix}\,,\label{eq:matrixM4}
\end{align}
\end{subequations}
with $M_4\in \mathbb{R}^{m\times m}$.
After setting $m=n$, differences between \cref{eq:findiff-2,eq:TLshortmatrixeq} are of subleading order in $n$.

\section{Impedance from TL equations}\label{sec:ZfromTLEq}
Having derived TL equations from their corresponding circuits in \cref{sec:circuit_to_ODE}, we now derive the pore impedances $Z_p$, $Z_{\rm con}$, and $Z_{F}$ from these TL equations.
We highlight the differences between our derivations and those found in the literature.

\subsection{$Z_p$ from TL equation for standard TL circuit}\label{sec:Zreflectivebc}
We start by considering a case without surface conduction ($R_p/R_F=0$), for which the TL equation [\cref{eq:TLeqfaradaic}] reduces to 
\begin{subequations}\label{eq:TLeqPosey}
\begin{align}
    R_p C  \partial_{t}\psi_d&=\ell_p^2 \partial_{z}^{2}\psi_d\,, &\quad &z\in[0,\ell_p]\,,\\
    \psi_d(z,0)&=0\,, \label{eq:TLinit2}\\
    \ell_p\partial_{z}\psi_d(0,t)&=\xi[\psi_d(0,t)-\Psi(t)], \label{eq:TLbcs_improved}\\
    \partial_{z}\psi_d(\ell_p,t)&=0\,.\label{eq:TLbcsNeumann}
\end{align}
\end{subequations}
In the case of impedance spectroscopy with no bias potential, the wall potential reads $\Psi(t)=\Psi_0\sin(\omega t)$.
By performing Laplace transformations [for a general function $f(t)$, we write $\hat{f}(s)\equiv \mathcal{L}\left\{f(t)\right\}\equiv\int_{0}^{\infty}\diff t f(t)\exp{(-ts)}$] and using $\mathcal{L}\left\{\partial_t f(x,t)\right\}=s\hat{f}(x,s)-f(x,0)$, we find
\begin{subequations}\label{eq:TLlaplace}
\begin{align}
    sR_p C \hat{\psi}_d&=\ell_p^2\partial_{z}^{2}\hat{\psi}_d\,, & &z\in[0,\ell_p]\,,\label{TLlaplace1}\\
    \ell_p\partial_{z}\hat{\psi}_d(0,s)&=\xi[\hat{\psi}_d(0,s)-\hat{\Psi}(s)], \label{TLlaplacebc3}\\
    \partial_{z}\hat{\psi}_d(\ell_p,s)&=0\,,\label{TLlaplacebc2}
\end{align}
\end{subequations}
whose solution reads
\begin{equation}\label{eq:psihatsolution}
    \hat{\psi}_d(z,s)=\frac{\hat{\Psi}(s)\cosh [\sqrt{sR_p C }(z/\ell_p-1)]}{\xi^{-1}\sqrt{sR_p C }\sinh\sqrt{sR_p C }+\cosh \sqrt{sR_p C }}\,.
\end{equation}
We can now find the current into the pore with $\hat{I}(s)=-\ell_p\partial_z \hat{\psi}_d(0,s)/R_p$ [see discussion below \cref{eq:currentcontinuation}], giving 
\begin{equation}\label{eq:currentcorrect}
    \hat{I}(s)=\frac{\hat{\Psi}(s)}{R_p} \frac{\sqrt{sR_p C }\sinh \sqrt{sR_p C }}{\xi^{-1}\sqrt{sR_p C }\sinh\sqrt{sR_p C }+\cosh \sqrt{sR_p C }}.
\end{equation}
This yields the impedance
\begin{equation}\label{eq:Ztotal}
    \hat{Z}(s)\equiv \frac{\hat{\Psi}(s)}{\hat{I}(s)}= R_{r}+R_p\frac{\coth \sqrt{sR_p C }}{\sqrt{sR_p C }}.
\end{equation}
Generally, the complex Laplace variable can be written as $s=\varsigma+\im \omega$. 
We set $\varsigma=0$ as we are interested in the steady state.
\Cref{eq:Ztotal} is then identical to \cref{eq:ZWO_total}. 

This derivation of $R_{r}+Z_p$ differs from the one found in the literature (both old \cite{levie1963} and recent \cite{lasia2014electrochemical,huang2020editors}) in one crucial point: the boundary condition \cref{eq:TLbcs_improved}.
We showed in \cref{sec:circuit_to_ODE} how \cref{eq:TLeqPosey} is equivalent to a matrix differential equation based on combining Ohm's and Kirchhoff's laws for all the nodes of the TL circuit.
Hence, the Robin boundary condition \cref{eq:TLbcs_improved} physically signals the conservation of ionic current. 
It is easier to see this if we rewrite \cref{eq:TLbcs_improved} in terms of the centerline potential, $\psi_c(z,t)=\Psi(t)-\psi_d(z,t)$, to 
\begin{equation}\label{eq:currentcontinuation}
    \frac{\psi_c(0,t)}{R_{r}}=\frac{\ell_p}{R_p}\partial_{z}\psi_c(0,t)\,.
\end{equation}
Here, the left-hand side gives the ionic current from the reservoir into the pore $(z=0^-)$.
As the TL model does not explicitly account for the reservoir at $z<0$, Ohm's law for this region is expressed in terms of the total potential drop over the reservoir [$\psi_c(0,t)$]. 
The right-hand side of \cref{eq:currentcontinuation} represents the ionic current in the pore at $z=0^+$.
A partial derivative appears here as the ionic current in the pore is driven by an electric field $-\partial_{z}\psi_c(z,t)$, which varies in the pore.
\Cref{eq:currentcontinuation} is thus a statement of current conservation.

Instead of \cref{eq:TLbcs_improved}, de Levie applied a Dirichlet boundary condition at the pore-reservoir interface \cite{levie1963}, $\psi_d(0,t)=\Psi(t)$.
As this corresponds to the $\xi\to\infty$ limit of \cref{eq:TLbcs_improved}, we immediate find $\hat{I}(s)=(\hat{\Psi}(s)/R_p) \sqrt{sR_p C }\tanh \sqrt{sR_p C }$ by taking $\xi\to\infty$ in \cref{eq:currentcorrect}.
The impedance of the pore then amounts to $\hat{Z}(s)= R_p\coth (\sqrt{sR_p C })/\sqrt{sR_p C }$, which is identical to $Z_p$.
In turn, the reservoir can be reintroduced by connecting $Z_p$ in series with $R_{r}$, yielding \cref{eq:Ztotal}. 
The problem with this derivation is that the Dirichlet boundary condition fixes the local potential drop at $z=0$, but one cannot enforce the potential there.
Experimentally, one controls the potential difference between the pore wall and some far-away counter (and reference) electrode.
Moreover, with the Dirichlet boundary condition, the physical interpretation of current conservation between the pore and the reservoir to which it is attached is lost.
Interestingly, even though the usual derivation of \cref{eq:Ztotal} used wrong boundary conditions for the pore-reservoir connection, fixing this error led to the same impedance, $Z_p$. 

\subsection{$Z_{\rm con}$ from TL equation for TL circuit with contact resistance}
The circuit with a contact resistance [\cref{TLcircuit_contact_resistance}] is governed by \cref{eq:TLeqshort} [different from \cref{eq:TLeqPosey} in the boundary condition at $z=\ell_p$], which is solved by
\begin{equation}
    \hat{\psi}_d(z,s)=- \frac{\hat{\Psi}(s)\sinh [\sqrt{sR_p C }(z/\ell_p-1)]}{\xi^{-1}\sqrt{sR_p C }\cosh\sqrt{sR_p C }+\sinh \sqrt{sR_p C }}\,,
\end{equation}
instead of \cref{eq:psihatsolution}.
Again, calculating the current by $\hat{I}(s)=-\ell_p\partial_z \hat{\psi}_d(0,s)/R_p$, we find the impedance 
\begin{equation}
    \hat{Z}(s)= R_{r}+R_p\frac{\tanh \sqrt{sR_p C }}{\sqrt{sR_p C }},
\end{equation}
in agreement with \cref{eq:ZWS_total}.

\subsection{Faradaic pore impedance $Z_{F}$ from TL equation for the leaky TL circuit}
The TL equation of the leaky TL circuit was stated in \cref{eq:TLeqfaradaic}. 
Tracing the steps we set in \cref{sec:Zreflectivebc}, we see that \cref{TLlaplace1} changes to
\begin{equation}\label{TLlaplacefaradaic}
    sR_p C \hat{\psi}_d =\ell_p^2 \partial_{z}^{2}\hat{\psi}_d-\frac{R_p}{R_F}\hat{\psi}_d\,, \qquad z\in[0,\ell_p]\,.
\end{equation}
By writing $s'=s+1/(R_F C)$, we find that \cref{eq:Ztotal} again holds, but with $s$ replaced by $s'$, which is identical to $R_{r}+Z_{F}$ [\cref{eq:ZWfaradaic_total}].

\section{Impedance from step response}\label{sec:stepresponse}
A system's impedance, $\hat{Z}(s)=\hat{\Psi}(s)/\hat{I}(s)$, is usually measured by subjecting it to a small-amplitude sinusoidal voltage. 
One can also find the same impedance using any other voltage perturbation as long as 1) it contains all frequencies, and 2) the perturbation is small \cite{pilla1970transient} (see also Yoo and Park~\cite{yoo2000electrochemical} and Sec. 3.7 of Lasia~\cite{lasia2014electrochemical}).
Hence, the impedance also follows from the current in response to a potential step $\Psi_{\rm step}(t)=\Psi_0\Theta(t)$, with $\Theta(t)$ being the Heaviside step function, as
\begin{equation}\label{eq:Zstep}
    \hat{Z}(s)=\frac{\mathcal{L}\left\{\Psi_{\rm step}(t)\right\}}{\mathcal{L}\left\{I_{\rm step}(t)\right\}}=\frac{\Psi_0}{s}\frac{1}{\mathcal{L}\left\{I_{\rm step}(t)\right\}}\,.
\end{equation}
In \cref{sec:numericalstudy}, we will use \cref{eq:Zstep} to numerically determine the impedance of a continuum pore model from its step response.
But first, we show how the TL model's impedance follows from its step response.

\subsection{TL equation step response}\label{sec:TLequationstep}
\begin{figure}
    \includegraphics[width=\linewidth]{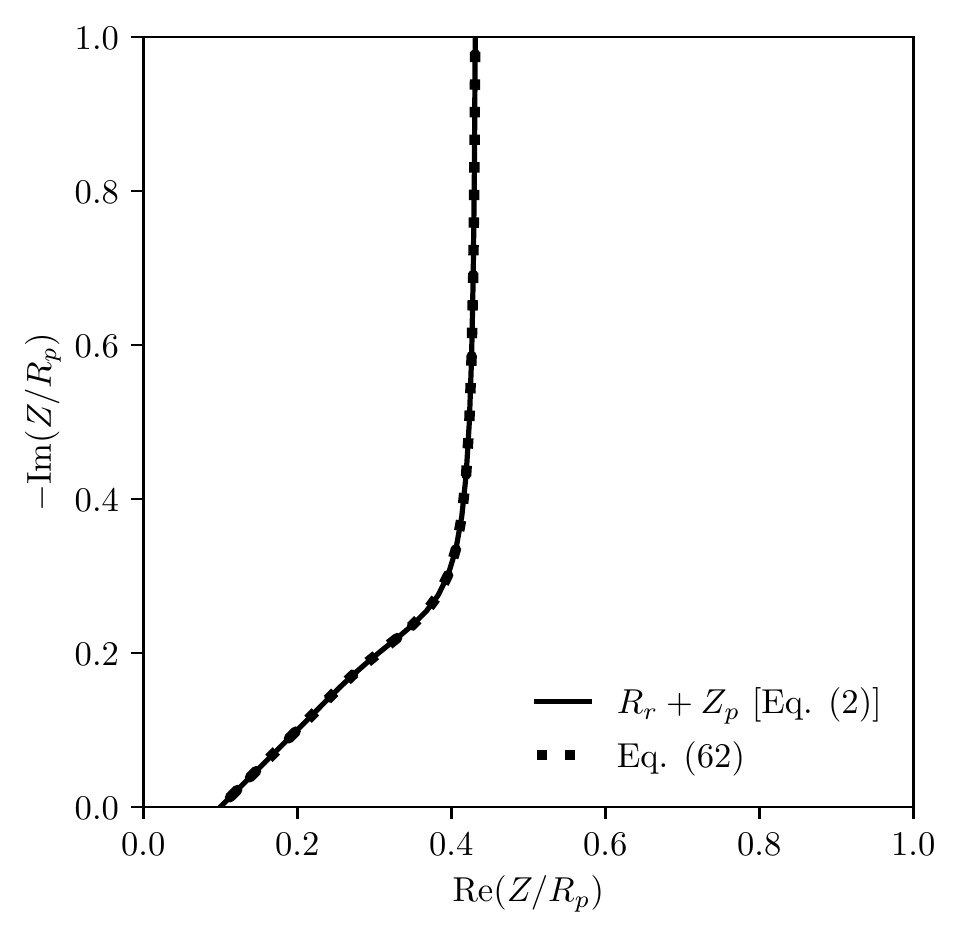}
    \caption{Plot of \cref{eq:ZWO_total,eq:Zfromstep} for $R_p/R_{r}=10$. \label{fig2}}
\end{figure}

Posey and Morozumi \cite{posey1966theory} solved \cref{eq:TLeqPosey} for the case $\Psi(t)=\Psi_{\rm step}(t)$ and found 
\begin{equation}\label{eq:poseyandmorozumi}
    \frac{\psi_c(z,t)}{\Psi(t)}=\sum_{j\ge1}\frac{4\sin \alpha_j \cos \left[\alpha_j\left(1-z/\ell_p\right)\right] }{2\alpha_j +\sin 2\alpha_j}\exp{\!\left(-\frac{\alpha_j^2 t}{R_p C }\right)}\,,
\end{equation}
where $\alpha_j$ with $j=1,2,\ldots$ are the solutions of the transcendental equation
\begin{equation}
    \alpha_j \tan \alpha_j = \xi\,.\label{eq:continuumtranscendental}
\end{equation}
The current $I(t)=\ell_p\partial_z \psi_c(0,s)/R_p$ into the pore amounts to
\begin{equation}\label{eq:PoseyandMorozumicurrent}
    I_{\rm step}(t)= \frac{\Psi_0}{R_p}\Theta(t)\sum_{j\ge1}\frac{4 \alpha_j \sin^2 \alpha_j}{2\alpha_j +\sin 2\alpha_j} \exp{\left(-\frac{\alpha_j^2 t}{R_p C }\right)}\,.
\end{equation}
Inserting \cref{eq:PoseyandMorozumicurrent} into \cref{eq:Zstep}, we find 
\begin{equation}\label{eq:Zfromstep}
    \frac{\hat{Z}(s)}{R_p}=\left( \sum_{j\ge1}\frac{4 \alpha_j \sin^2 \alpha_j}{2\alpha_j +\sin 2\alpha_j} \frac{sR_p C  }{\alpha_j^2 +sR_p C }\right)^{-1}\,.
\end{equation}
While it is not clear how \cref{eq:Zfromstep} relates to $R_{r}+Z_p$ [\cref{eq:ZWO_total}], \cref{fig2} shows that they overlap.
This overlap can be understood for the case $\xi=R_p/R_{r}\to\infty$, when $\alpha_j=(j-1/2)\pi$ solves \cref{eq:continuumtranscendental}, and \cref{eq:Zfromstep} simplifies to
\begin{equation}\label{eq:ZWOnoreservoir}
    \frac{\hat{Z}(s)}{R_p}=\left( \sum_{j\ge1}\frac{2 sR_p C  }{(j-1/2)^2\pi^2 +sR_p C }\right)^{-1}\,.
\end{equation}
Now, inserting the Weierstrass factorization of the hyperbolic cosine with complex argument \cite{montella2020voigt}
\begin{equation}
    \cosh (z)=\prod_{j=1}^\infty \left(1+\frac{z^2}{(j-1/2)^2\pi^2}\right)
\end{equation}
into the right-hand side of 
\begin{equation}
    \frac{\coth z}{z}=\frac{1}{z}\left(\frac{\partial \ln \cosh z}{\partial z}\right)^{-1}
\end{equation}
yields
\begin{equation}
    \frac{\coth z}{z}=\left(\sum_{j\ge1} \frac{2z^2}{(j-1/2)^2\pi^2+z^2}\right)^{-1}\,.
\end{equation}
For $z=\sqrt{\im \omega R_p C }$, we then recover $Z_p$ on the left-hand side and \cref{eq:ZWOnoreservoir} on the right-hand side.

\subsection{Overlapping EDLs}
The regular TL equation describes the charging of a pore whose EDLs are much thinner than the pore radius, $\lambda_D\ll\varrho_p$.
Henrique, Zuk, and Gupta studied the charging of pores with an arbitrary EDL thickness \cite{henrique2021charging}.
Specifically, they analytically solved the PNP equations [viz. \cref{eq:numPNP1a}] for a cylindrical pore subject to a small applied potential, for which they found the centerline potential
\begin{align}\label{eq:henriquezukgupta}
    &\frac{\psi_c(z,t)}{\Psi(t)}=I_{0}\left(\frac{\varrho_p}{\lambda_D}\right)^{-1}+\left[1-I_{0}\left(\frac{\varrho_p}{\lambda_D}\right)^{-1}\right]\times\nn
    &\times\sum_{j\ge1}\frac{4\sin \alpha_j \cos \left[\alpha_j\left(1-z/\ell_p\right)\right] }{2\alpha_j +\sin 2\alpha_j}
    \exp{\!\left(-\frac{\alpha_j^2 t}{\tau}\right)}\,,
\end{align}
where $\tau=2\ell_p^2\lambda_D/(\varrho_p D)\times I_{1}\left(\varrho_p/\lambda_D\right)/I_{0}\left(\varrho_p/\lambda_D\right)$ and where $\alpha_j$ with $j=1,2,\ldots$ are the solutions of the transcendental equation
\begin{equation}
    \alpha_j \tan \alpha_j = \frac{\ell_p}{\ell_r}\frac{\varrho_r^2}{\varrho_p^2}\,.\label{eq:continuumtranscendental2}
\end{equation}
\Cref{eq:henriquezukgupta} does not relax to $\psi_c(z,t)=0$ at late times when $\varrho_p/\lambda_D\sim1$.
Therefore, when Henrique, Zuk, and Gupta interpreted their PNP model in terms of a TL-like ladder circuit, they had to include an interfacial resistance that grew monotonously over time \cite{henrique2021charging}.
Moreover, they argued that the conductivity of the electrolyte in the pore changes from $\kappa$ to $\kappa_H=\kappa I_{0}\left(\varrho_p/\lambda_D\right)/[I_{0}\left(\varrho_p/\lambda_D\right)-1]$.
Hence, while the right-hand side of \cref{eq:continuumtranscendental2} looks like the ratio of the pore to reservoir resistance, that interpretation only holds for $\varrho_p\gg\lambda_D$, as the Bessel function factor in $\kappa_H$ then tends to unity. 
(In \cit{henrique2021charging}, the right-hand side of \cref{eq:continuumtranscendental2} also contained the ratio reservoir to pore diffusivities, which we consider here to be unity.)

Retracing our steps of \cref{sec:TLequationstep}, we now find 
\begin{align}\label{eq:ZfromstepGupta}
    \frac{\hat{Z}(s)}{R_p}&=
    \left(\sum_{j\ge1}\frac{4 \alpha_j \sin^2 \alpha_j}{2\alpha_j +\sin 2\alpha_j} \frac{s\tau }{\alpha_j^2 +s\tau}\right)^{-1}\,,
\end{align}
where we absorbed a factor $I_{0}\left(\varrho_p/\lambda_D\right)/[I_{0}\left(\varrho_p/\lambda_D\right)-1]$ into $R_p$. 
Apart from that factor, we see that EDL overlap leads to a shift of frequencies (through $\tau$) as compared to \cref{eq:Zfromstep}.

\section{Numerical study of pore charging}\label{sec:numericalstudy}
\subsection{Setup}
We delineate the validity of the pore impedance $Z_p$ for pores of various aspect ratios by numerical simulations of their charging.
As we are interested in a pore's impedance, its response to a small amplitude voltage, we ignore fluid flow, as the electroconvective term in the Navier Stokes equations is quadratic (and thus subleading) in the electric field \cite{malgaretti2019driving}.
We first discuss numerical PNP simulations of pore charging in response to applied step potentials, much like Yang and coworkers \cite{yang2022simulating}.
From these data, we determine the corresponding impedance $\hat{Z}(s)$ using the method of \cref{sec:stepresponse}.

\begin{figure}
    \includegraphics[width=\linewidth]{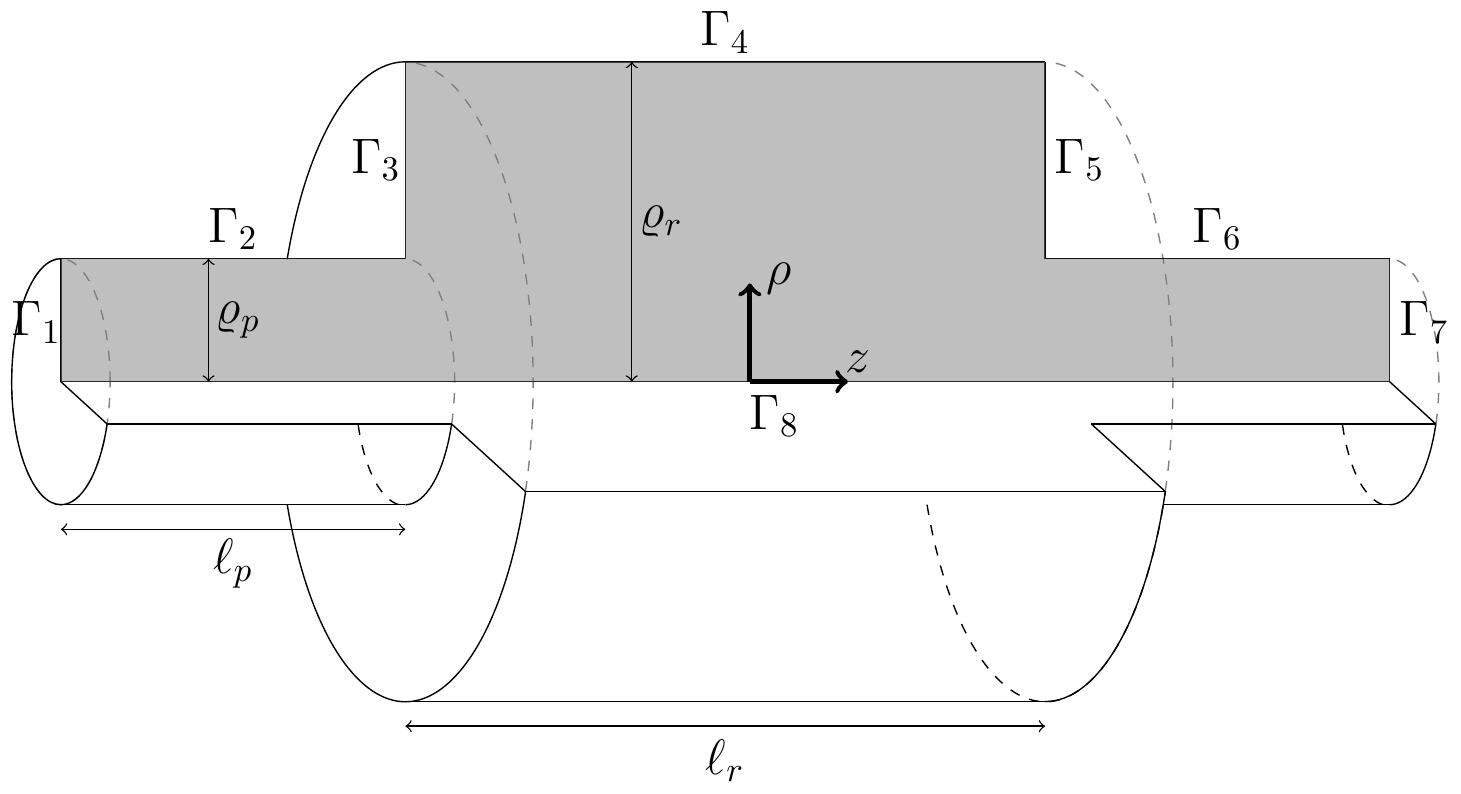}
    \caption{Schematic (not to scale) of an axisymmetric supercapacitor model consisting of two pores (left and right) connected to a reservoir (middle). Our two-dimensional numerical domain is colored grey, with the coordinate system's origin set to the right pore's entrance.}
    \label{fig:schematic}
\end{figure}

\Cref{fig:schematic} shows our system of interest. 
We consider two cylindrical pores of radius $\varrho_p$ and length $\ell_p$ connected on either side of a cylindrical reservoir of radius $\varrho_r$ and length $\ell_r$; all the cylinders' axes are aligned, so the whole system is axisymmetric. 
We use a cylindrical coordinate system $\boldsymbol{r}=(\rho, z, \theta)$ with $\rho$ and $z$ being the radial and longitudinal coordinates, respectively.
We set $z=0$ at the entrance of the right pore to make comparisons to the TL model easier, as that model only explicitly treats the pore, with the effect of the reservoir captured in the boundary condition at $z=0$.

We model the spatiotemporal evolution of the local electrostatic potential $\psi(\boldsymbol{r},t)$ and the local cationic and anionic densities $c_\pm(\boldsymbol{r},t)$ in our setup through the PNP equations, 
\begin{subequations}\label{eq:numPNP1a}
\begin{align}
    \varepsilon \nabla^2\psi &= -e(c_+ - c_-),\\
    \partial_t c_{\pm} &= -\boldsymbol{\nabla}\cdot \boldsymbol{j}_{\pm},\label{eq:numPNP2a}\\
    \boldsymbol{j}_{\pm} &= -D\left(\boldsymbol{\nabla} c_{\pm} \pm c_{\pm}\beta e\boldsymbol{\nabla}\psi\right),
\end{align}
\end{subequations}
where $\varepsilon$ is the electrolytic permittivity, $e$ is the unit charge, $D$ is the ionic diffusion coefficient (taken equal among cations and anions), and $\boldsymbol{j}_\pm$ are the cationic and anionic fluxes.
Moreover, $\beta=1/(k_B T)$ is the inverse thermal energy, where $k_B$ is Boltzmann's constant and $T$ is the temperature.

We consider all pore and reservoir walls blocking and set the initial ionic densities to $c_0$ throughout the system. 
At the time $t=0$, we apply a potential difference $2\Psi_0$ between the pores, which, due to the symmetry of our setup, is shared evenly between the pores.
The following initial and boundary conditions thus apply
\begin{subequations}\label{eq:numBCs}
\begin{align}
    c_{\pm}(\boldsymbol{r}, t=0) &= c_0,\\
    -\psi\big |_{\Gamma_2} = \psi\big |_{\Gamma_6} &= \Psi_0,\label{eq:numBCs_potential}\\
    \boldsymbol{\nabla}\psi\cdot\boldsymbol{n}\big |_{\Gamma_1,\Gamma_3, \Gamma_4,\Gamma_5,\Gamma_7,\Gamma_8} &= 0,\label{eq:numBCs_nocharge}\\
    \boldsymbol{j}_{\pm}\cdot\boldsymbol{n}\big |_{\Gamma_1,\Gamma_2, \Gamma_3, \Gamma_4, \Gamma_5,\Gamma_6,\Gamma_7,\Gamma_8} &= 0,\label{eq:numBCs_noflux}
\end{align}
\end{subequations}
with $\boldsymbol{n}$ being the outwards pointing normal vector at each boundary.
\Cref{eq:numBCs_nocharge} says that the respective boundaries are uncharged, which applies to dielectric materials.
In \cref{sec:conductingends}, we discuss a case where the boundaries $\Gamma_1$ and $\Gamma_7$ are conducting instead.

\begin{figure*}
    \includegraphics[width=\linewidth]{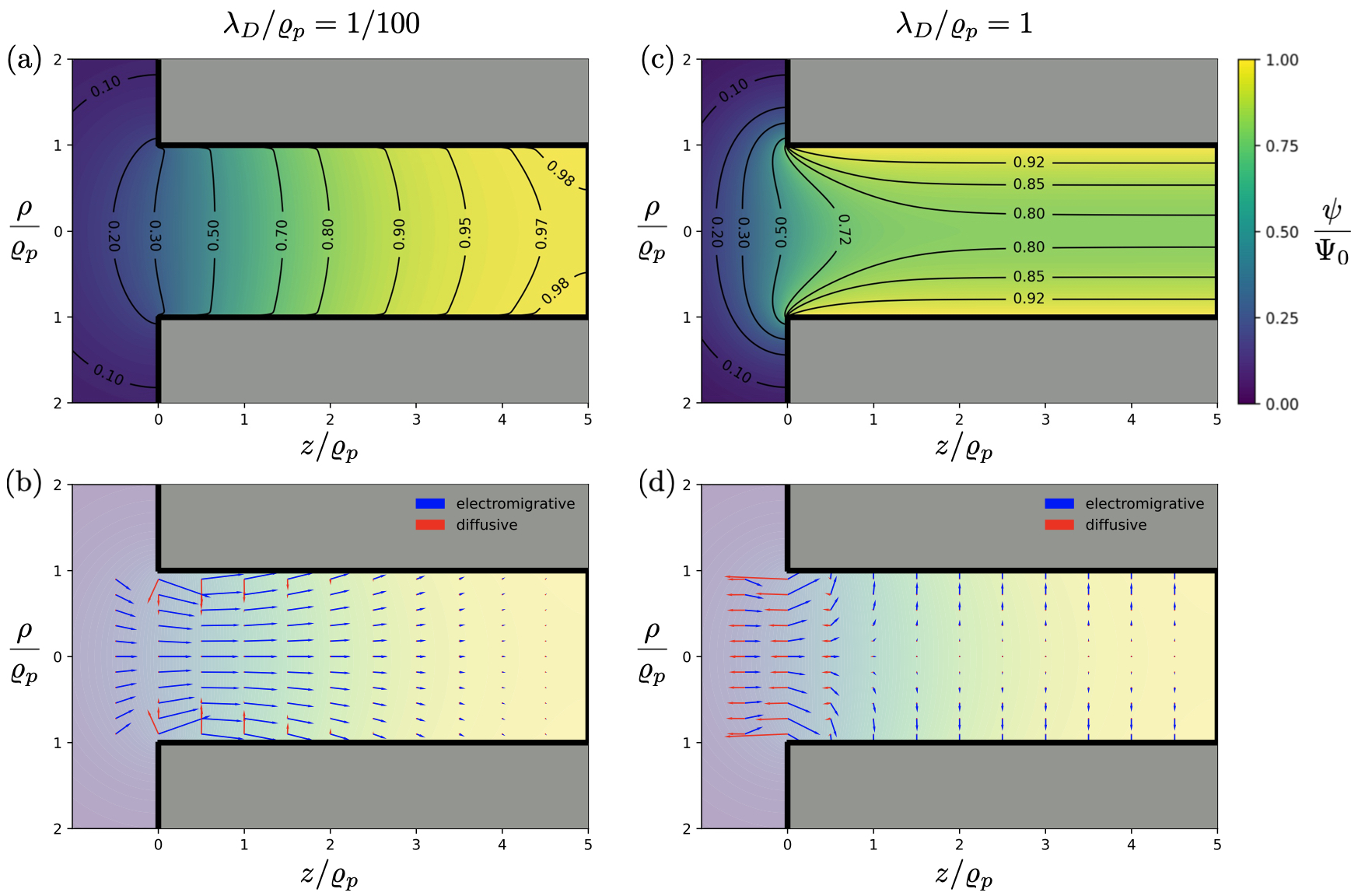}
    \caption{Snapshot of numerical PNP solutions (a,c) and the ionic flux contributions (b,d) for a pore-reservoir system for $\ell_r=20\varrho_p$, $\varrho_r=10\varrho_p$, $\ell_p=5\varrho_p$ and $\lambda_D/\varrho_p=1/100$ at $\bar{t}=0.039$ (a,b) and $\lambda_D/\varrho_p=1$ at $\bar{t}=52$ (c,d). In panel (b), the diffusive flux vector is scaled 100 times larger than the electromigrative flux vector. Only one pore and the immediate vicinity of the reservoir are shown here.}
    \label{fig:potential}
\end{figure*}

In our axisymmetric setup, all $\theta$ dependence drops, so that $\psi=\psi(\rho,z,t)$, $c_\pm=c_\pm(\rho,z,t)$, $\boldsymbol{j}_{\pm}=j_{\rho,\pm}(\rho,z,t)\hat{\boldsymbol{\rho}}+j_{z,\pm}(\rho,z,t)\hat{\mathbf{z}}$ (in this section alone, symbols with hats refer to unit vectors, not Laplace transformed variables) and \cref{eq:numPNP1a} reads 
\begin{subequations}\label{eq:numPNP1}
\begin{align}
     \rho^{-1}\partial_{\rho}(\rho\partial_{\rho}\psi) + \partial_z^2\psi &= -\frac{e}{\varepsilon}(c_+ - c_-),\\
    \partial_t c_{\pm} &= - \rho^{-1}\partial_{\rho}(\rho j_{\rho,\pm}) - \partial_z j_{z,\pm},\label{eq:numPNP2}\\
    j_{\rho,\pm} &= -D\left(\partial_{\rho} c_{\pm} \pm c_{\pm}\beta e\partial_{\rho}\psi\right),\\
    j_{z,\pm} &= -D\left(\partial_z c_{\pm} \pm c_{\pm}\beta e\partial_z\psi\right).
\end{align}
\end{subequations}
In our setup, the normal vector amounts to $\boldsymbol{n}=\hat{\boldsymbol{\rho}}$ on $\Gamma_2,\Gamma_4$, and $\Gamma_6$, to $\boldsymbol{n}=-\hat{\boldsymbol{\rho}}$ on $\Gamma_8$, to $\boldsymbol{n}=-\hat{\mathbf{z}}$ on $\Gamma_1$ and $\Gamma_3$ and to $\boldsymbol{n}=\hat{\mathbf{z}}$ on $\Gamma_5$ and $\Gamma_7$.
Hence, \cref{eq:numBCs_nocharge,eq:numBCs_noflux} amount to 
\begin{subequations}\label{eq:numBCs2}
\begin{align}
    \partial_{\rho}\psi |_{\Gamma_2,\Gamma_4,\Gamma_6,\Gamma_8}=
    \partial_{z}\psi |_{\Gamma_1,\Gamma_3,\Gamma_5,\Gamma_7}&=0,\\
    j_{\rho,\pm}\big |_{\Gamma_2,\Gamma_4,\Gamma_6,\Gamma_8} =
    j_{z,\pm}\big |_{\Gamma_1,\Gamma_3,\Gamma_5,\Gamma_7}&= 0.
\end{align}
\end{subequations} 

We scale all lengths by the pore radius: $\bar{z}=z/\varrho_p$ and $\bar{\rho}=\rho/\varrho_p$, with the bar notation indicating dimensionless quantities. 
We also use the dimensionless time $\bar{t}=Dt/\varrho_p^2$, potential $\bar{\psi}=\beta e\psi$, ion densities $\bar{c}_\pm=c_\pm/c_0$, and fluxes $\bar{\boldsymbol{j}}_\pm=\boldsymbol{j}_\pm \varrho_p/(Dc_0)$.
When inserted into \cref{eq:numPNP1,eq:numBCs,eq:numBCs2}, we obtain the dimensionless PNP equations,
\begin{subequations}\label{eq:nondim-numPNP}
\begin{align}
    \bar{\rho}^{-1}\partial_{\bar{\rho}}(\bar{\rho}\partial_{\bar{\rho}}\bar{\psi}) + \partial_{\bar{z}}^2\bar{\psi} &= -\frac{1}{2}\frac{\varrho_p^2}{\lambda_D ^2}(\bar{c}_+ - \bar{c}_-),\label{eq:nondim-numPNP1}\\
    \partial_{\bar{t}}\bar{c}_{\pm} &= - \bar{\rho}^{-1}\partial_{\bar{\rho}}(\bar{\rho}j_{\bar{\rho},\pm}) - \partial_{\bar{z}} j_{\bar{z},\pm},\label{eq:nondim-numPNP2}\\
    \bar{j}_{\bar{\rho},\pm} &= -\partial_{\bar{\rho}}\bar{c}_{\pm} \mp\bar{c}_{\pm}\partial_{\bar{\rho}}\bar{\psi},\\
    \bar{j}_{\bar{z},\pm} &= -\partial_{\bar{z}}\bar{c}_{\pm} \mp \bar{c}_{\pm}\partial_{\bar{z}}\bar{\psi},
\end{align}
\end{subequations}
and associated initial and boundary conditions 
\begin{subequations}\label{eq:PNPbcsdimless}
\begin{align}
    \bar{c}_{\pm}(\bar{\rho}, \bar{z}, \bar{t}=0) &= 1,\\
    -\bar{\psi}\big |_{\Gamma_2} = \bar{\psi}\big |_{\Gamma_6} &= \bar{\Psi}_0,\label{eq:appliedpotential}\\
    \partial_{\bar{\rho}}\bar{\psi} |_{\Gamma_4,\Gamma_8}=\partial_{\bar{z}}\bar{\psi} |_{\Gamma_1,\Gamma_3,\Gamma_5,\Gamma_7}&=0,\label{eq:nonchargedboundary}\\
    \bar{j}_{\bar{\rho},\pm}\big |_{\Gamma_2,\Gamma_4,\Gamma_6,\Gamma_8} = \bar{j}_{\bar{z},\pm}\big |_{\Gamma_1,\Gamma_3,\Gamma_5,\Gamma_7}&= 0.
\end{align}
\end{subequations}
In \cref{eq:nondim-numPNP1}, $\lambda_D = 1/\sqrt{8\pi \lambda_B c_0}$ is the Debye length (the characteristic width of the EDL), where $\lambda_B=\beta e^2/(4\pi \varepsilon )$ is the Bjerrum length.

We will solve \cref{eq:nondim-numPNP1,eq:nondim-numPNP2} by the finite element method (FEM).
To do so, we multiply them with test functions $v$ and $q$, respectively, integrate over the domain and apply the boundary conditions \cref{eq:PNPbcsdimless}. 
This yields their variational formulation
\begin{align}\label{eq:variational1}
    &\int_{\Gamma_2,\Gamma_6}\partial_{\bar{\rho}}\bar{\psi} v \diff \bar{z} - \int_{\Omega}\partial_{\bar{\rho}}\bar{\psi}\partial_{\bar{\rho}} v \bar{\rho} \diff \bar{\rho}\diff \bar{z} -\int_{\Omega}\partial_{\bar{z}}\bar{\psi}\partial_{\bar{z}} v \bar{\rho}\diff \bar{\rho}\diff \bar{z} \nn
    &\qquad=-\frac{1}{2}\frac{\varrho_p^2}{\lambda_D ^2}\int_{\Omega}(\bar{c}_+ - {\bar{c}_-})v\bar{\rho}\diff \bar{\rho}\diff \bar{z} 
\end{align}
and
\begin{align}\label{eq:variational2}
    \int_{\Omega} q\partial_{\bar{t}}\bar{c}_{\pm}\,\bar{\rho}\diff \bar{\rho}\diff \bar{z} 
    &= \int_{\Omega} \bar{j}_{\bar{\rho},\pm}\partial_{\bar{\rho}} q \,\bar{\rho}\diff \bar{\rho}\diff \bar{z}\nn 
    &\qquad\qquad + \int_{\Omega}\bar{j}_{\bar{z},\pm}\partial_{\bar{z}} q\,\bar{\rho}\diff \bar{\rho}\diff \bar{z} \,.
\end{align}
We discretized \cref{eq:variational1,eq:variational2} using linear elements and solved them implicitly and coupled using a Newton solver from the FEniCS library \cite{logg2012automated}. 
The mesh is generated using Gmsh \cite{geuzaine2009gmsh}, with the spatial resolution at the pore wall being $0.001\varrho_p$, resolving the Debye length by at least 10 grid points. 
The solver code and the script to generate the mesh are on this \href{https://github.com/christian-pedersen/FEM-Axisymmetric-Poisson-Nernst-Planck/blob/main/README.md}{GitHub repository}.

We will study pore-reservoir systems with a fixed reservoir size of $\ell_r/\varrho_p = 20$ and $\varrho_r/\varrho_p=10$ and pore lengths of $\ell_p/\varrho_p=1, 2.5, 5, 10$, and $25$.
The ratio $\lambda_D / \varrho_p$ represents the EDL overlap---we will consider cases for which the EDLs are thin ($\lambda_D / \varrho_p=0.01$) and overlapping ($\lambda_D / \varrho_p=1$).
The dimensionless applied potential is set to $\bar{\Psi}_0=0.1$ throughout, corresponding to about \SI{2.5}{\milli\volt} for systems at room temperature.

\subsection{Step response}\label{sec:step_response_numerically}
\Cref{fig:potential}(a) and (c) show numerical solutions to \cref{eq:nondim-numPNP,eq:PNPbcsdimless} for the local potential $\psi(\rho,z,t)$ in a pore-reservoir system with $\lambda_D/\varrho_p=1/100$ at $\bar{t}=0.039$ (a) and with $\lambda_D/\varrho_p=1$ at $\bar{t}=52$ (c).
The figure shows that isopotential lines are not parallel to the pore's wall close to its entrance and end.
At this intermediate time, the pore has attracted counterions and developed EDLs near the reservoir-pore interface.
\Cref{fig:potential}(b) and (d) correspond to the same parameters as panels (a) and (c), respectively, and show the diffusive (red arrows) and electromigrative (blue arrows) contribution to the ionic fluxes, that is, the first and second terms on the right-hand side of
$\bar{j}_{\bar{\rho},+}-\bar{j}_{\bar{\rho},-} = -\partial_{\bar{\rho}}(\bar{c}_{+}-\bar{c}_{+}) -(\bar{c}_{+}+\bar{c}_{-})\partial_{\bar{\rho}}\bar{\psi}$.
Note that only in panel (b), corresponding to the same early time as in panel (a), we stretched the red arrows a hundredfold to make them visible compared to the blue arrows.
Hence, the ionic fluxes are almost entirely caused by the electric field, not by diffusion.
In panel (d), corresponding to the same parameters as panel (c) (overlapping EDLs and a late time), the strongest diffusive and electromigrative fluxes are near the pore-reservoir interface, where they nearly balance each other. 

\begin{figure}
    \includegraphics[width=\linewidth]{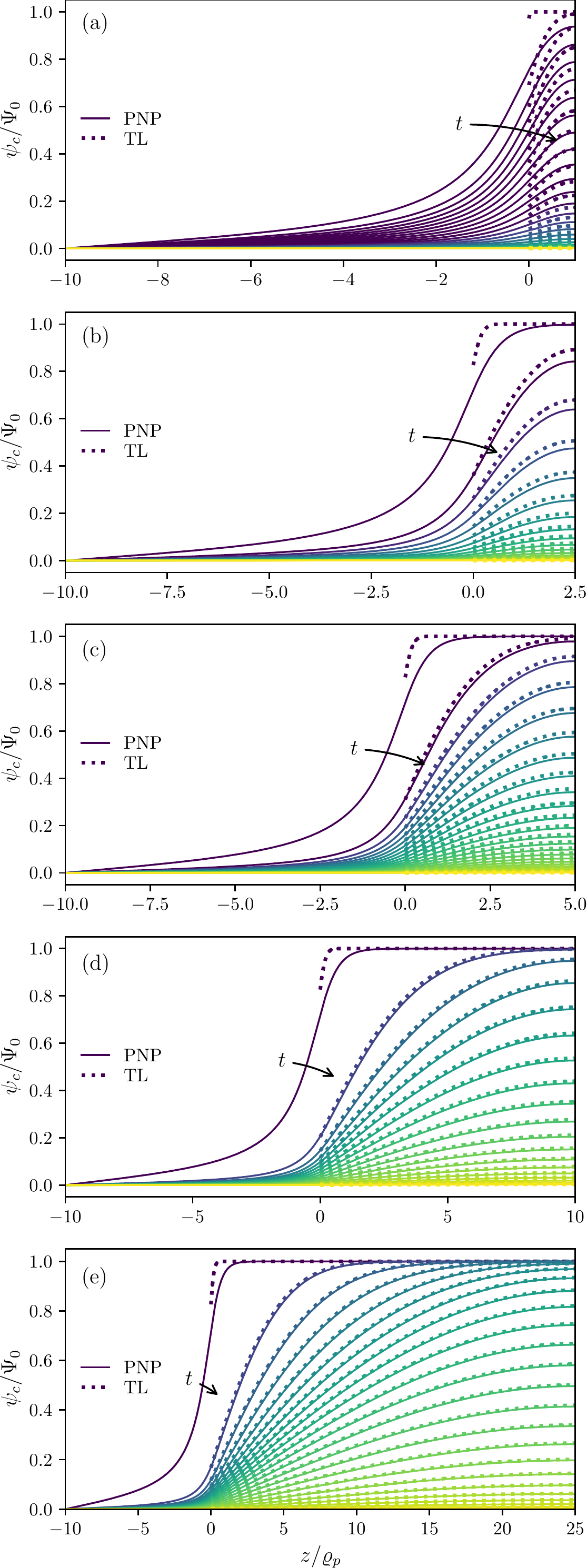}
    \caption{A pore's centerline potential from PNP (lines) and the TL model \cref{eq:poseyandmorozumi} (dotted) for a case with thin EDLs. 
    We set $\lambda_D/\varrho_p=1/100$ and $\ell_r=20\varrho_p$, $\varrho_r=10\varrho_p$ and, from top to bottom, $\ell_p=(1, 2.5, 5, 10, 25)\varrho_p$. 
    Colors in all panels refer to the same times in units of $\varrho_p^2/D$.}
    \label{fig:step_response}
\end{figure}

\begin{figure}
    \includegraphics[width=\linewidth]{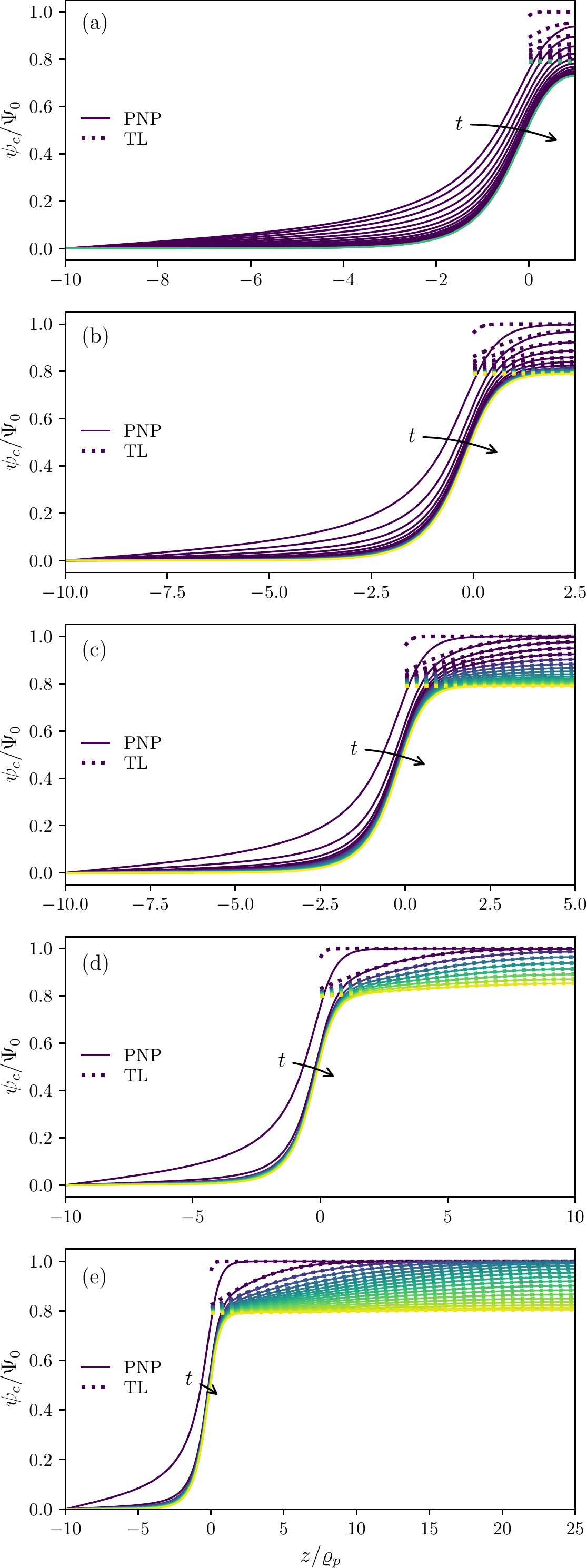}
    \caption{Same as \cref{fig:step_response}, except we consider overlapping EDLs, $\lambda_D/\varrho_p=1$, and plot the extended-TL model result \cref{eq:henriquezukgupta} instead of \cref{eq:poseyandmorozumi}.}
    \label{fig:step_response_Henrique}
\end{figure}

From the local potential $\psi(\rho,z,t)$, we find a pore's centerline potential and potential drop, previously studied through the TL model, by $\psi_c(z,t)=\psi(\rho=0,z,t)$ and $\psi_d(z,t)=\Psi-\psi(\rho=0,z,t)$, respectively. 
\Cref{fig:step_response} shows FEM solutions (lines) for $\psi_c(z,t)$ for various times, thin EDLs $\varrho_p/\lambda_D=100$, and various pore lengths in the different panels.
The colors in all panels refer to the same times in units of $\varrho_p^2/D$, where we picked colors from a purple to yellow scheme spanning the longest pore's ($\ell_p/\varrho_p=25$) relaxation. The shorter pores relax faster, so they are more purple.
\Cref{fig:step_response} also shows Posey and Morozumi's TL model solution \cref{eq:poseyandmorozumi}.
As expected, discrepancies between both methods are most apparent for short pores.
To draw \cref{eq:poseyandmorozumi}, we needed to specify $R_p/R_{r}$ for the different geometries.
We approximated the pore's resistance $R_p$ and (half) the reservoir's resistance $R_{r}$ by 
\begin{subequations}\label{eq:idealizedR}
\begin{align}
    R_p&=\frac{\ell_p}{\kappa \pi \varrho_p^2},\label{eq:Rpore}\\
    R_{r}&=\frac{\ell_r}{2\kappa \pi \varrho_r^2}+\frac{1}{4\kappa \varrho_p},\label{eq:Rreservoir}
\end{align}
\end{subequations}
where $\kappa$ is the electrolyte conductivity.
The first term in $R_{r}$ is the resistance of a cylindrical resistor between two flat plates; this term is the exact resistance for cases where $\varrho_r=\varrho_p$.
The second term in $R_{r}$ is Newman's resistance between a conducting disk and an infinitely large hemispherical electrode  \cite{newman1966resistance}.
The same resistance was later found by Hall, who identified it as the entrance resistance for ions entering a pore from a semi-infinite reservoir \cite{hall1975access}.
By approximating $R_r$ by the two terms in \cref{eq:Rreservoir}, we ensure we properly capture the reservoir resistance in the opposite limits of narrow and wide reservoirs.

\begin{figure*}
\includegraphics[width=\linewidth]{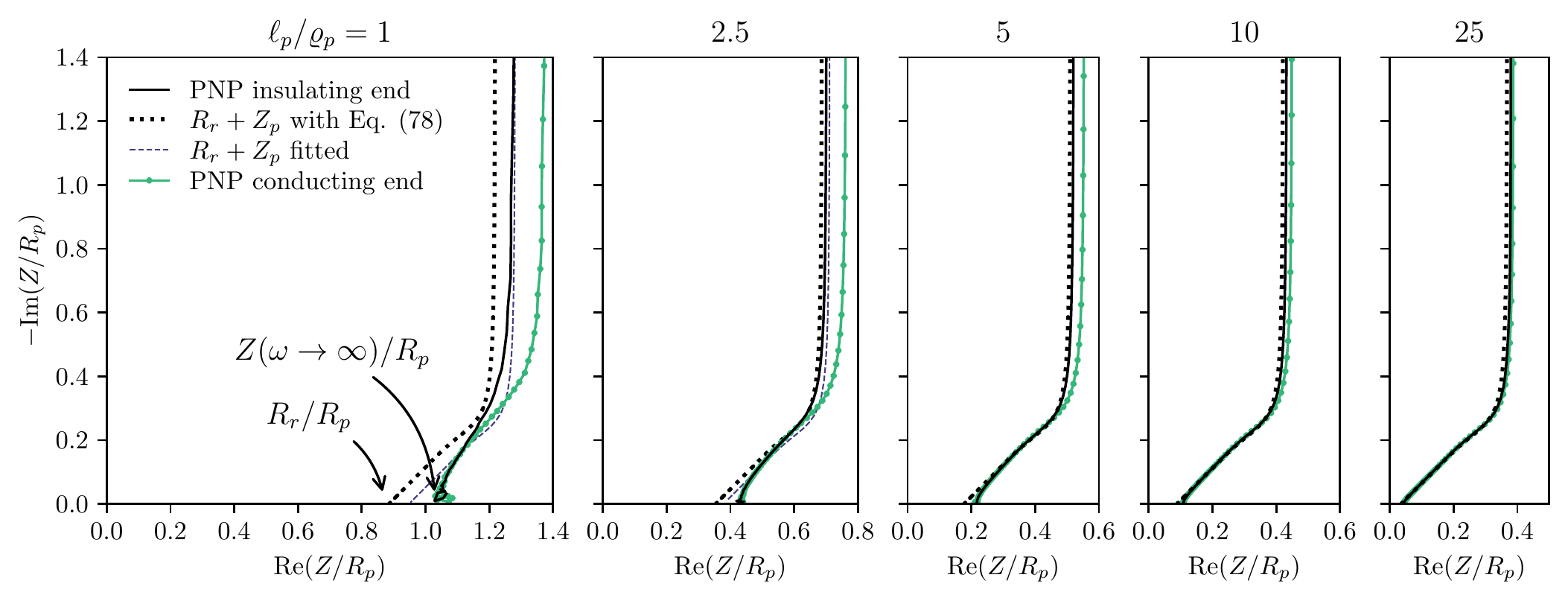}
    \caption{
    Impedance $Z$ of pores of different aspect ratios $\ell_p/\varrho_p$ with thin EDLs ($\lambda_D/\varrho_p=0.01$) and other parameters as in \cref{fig:step_response}. 
    The data corresponds to \cref{eq:ZWO_total} (dotted and dashed) and PNP step voltage solutions for insulating (black lines) and conducting (green lines with circles) pore ends.
    The pore impedance [\cref{eq:ZWO_total}] is scaled to the pore resistance $R_p$; the PNP data is scaled to $\ell_p/(\kappa\pi \varrho_p^2)$, i.e., the resistance of an isolated cylindrical electrolyte-filled pore.}
    \label{fig:impedance}
\end{figure*}

Yang and coworkers~\cite{yang2022simulating} also studied the charging of a pore in response to a step potential through the PNP equations but did not incorporate the Newman-Hall term in $R_{r}$.
That article noted that \cref{eq:poseyandmorozumi} does not capture a pore's early-time charging, especially near the pore-reservoir interface.
We found that adding the Newman-Hall resistance to $R_{r}$ yields better agreement between FEM solutions and \cref{eq:poseyandmorozumi}, even at early times; see \cref{{fig:step_response}}(d) and (e). 
Still, our expression for $R_{r}$ is an ad hoc combination of resistance expressions. 
The discrepancies that are still visible between both methods may be further reduced by using a better expression for $R_{r}$ and $R_p$.
Nevertheless, the impedance results discussed below [viz.~\cref{fig:impedance}] suggest that the TL model will never entirely capture the centerline potential's relaxation, even if one would have exact expressions for $R_{r}$ and $R_p$.

\Cref{fig:step_response_Henrique} is the same as \cref{fig:step_response} except for a different EDL overlap, $\varrho_p/\lambda_D=1$.
We now compare the FEM simulations of the PNP equations (lines) to \cref{eq:henriquezukgupta} (dotted lines).
To account for the Newman-Hall entrance resistance, we replaced the right-hand side of \cref{eq:continuumtranscendental2} with the ratio of \cref{eq:Rpore,eq:Rreservoir}.
Different from the case of thin EDLs, for overlapping EDLs, the late-time centerline potential transitions between $-4\sim z/\varrho_p\sim 2$ from a small value in the reservoir to a finite value in the pore.
\Cref{eq:henriquezukgupta} predicts that value to be $1/I_{0}\left(\varrho_p/\lambda_D\right)$, which amounts to 0.79 for $\varrho_p/\lambda_D=1$ as considered here.
As for pores with thin EDLs, for long pores with thick EDLs, the FEM solutions and \cref{eq:henriquezukgupta} agree decently.
For shorter pores, we see that \cref{eq:henriquezukgupta} overestimates the centerline potential.
The transition region between $-4\sim z/\varrho_p\sim 2$---visible for all pores---is not resolved by \cref{eq:henriquezukgupta}.

\subsection{Impedance}
\subsubsection{Numerical method}
To calculate the impedance from the step response data, we modify \cref{eq:Zstep} to
\begin{equation}\label{eq:znumerical}
    \hat{Z}(s)=\frac{\Psi_0}{\im \omega}\frac{1}{\mathcal{L}^{\text{num}}\left\{I_{\rm step}(t)\right\}}\,,
\end{equation}
where $\mathcal{L}^{\text{num}}$ is a numerical realization of the Laplace transform defined by
\begin{equation}\label{eq:Laplace-num}
    \mathcal{L}^{\text{num}}\big\{I(t)\big\}=\int_0^{t_{\text{max}}}I(t)\me^{-\im \omega t}\diff t,
\end{equation}
where $t_{\text{max}}$ is the last time of our numerical simulations. 
Integrating \cref{eq:Laplace-num} by parts and using $I=\diff Q/\diff t$, we find 
\begin{equation}\label{eq:Laplace-num-property}
    \mathcal{L}^{\text{num}}\big\{I(t)\big\} = Q(t_\text{max})\me^{-\im \omega t_\text{max}} - Q(0)+\im \omega \mathcal{L}^{\text{num}}\big\{Q(t)\big\}\,.
\end{equation}
As $t_\text{max}\to \infty$, the first term on the right-hand side drops and \cref{eq:Laplace-num-property} reduces to a known Laplace transform identity.

To determine $Q(t)$ from our $\bar{\psi}(\bar{z},\bar{\rho},\bar{t})$ data, we note that Gauss's law gives access to the boundary condition between a charged conductor next to an insulator, $e\sigma=- \varepsilon \boldsymbol{n}\cdot\mathbf{E}$, with $\sigma$ (\si{\per\meter\squared}) being the surface charge number density, $\mathbf{E}=-\boldsymbol{\nabla} \psi$ the local electric field, and $\boldsymbol{n}$ the normal vector into the conductor.
We have $\boldsymbol{n}=\hat{\boldsymbol{\rho}}$ on $\Gamma_6$, so $e\sigma= \varepsilon \partial_\rho \psi\big|_{\Gamma_6}$ or, in terms of the dimensionless potential,
\begin{equation}\label{eq:surfacechargedensity}
    \sigma = \frac{1}{4\pi\lambda_B}\partial_\rho \bar{\psi}\big|_{\Gamma_6}\,.
\end{equation}
The total charge on one pore $Q=e\int_{\Gamma_6}\diff A\, \sigma$ is thus
\begin{equation}\label{eq:Qnumerical}
    Q=\frac{\varrho_p e}{4\pi\lambda_B}2\pi\int_{0}^{\ell_p/\varrho_p}\diff \bar{z} \,\partial_{\bar{\rho}} \bar{\psi}(\bar{\rho}=1,\bar{z},\bar{t})\,.
\end{equation}
Our numerical solutions to the PNP equations give access to the dimensionless integral $\bar{Q}=2\pi\int_{0}^{\ell_p/\varrho_p}\diff \bar{z} \,\partial_{\bar{\rho}} \bar{\psi}(\bar{\rho}=1,\bar{t})$.
Putting \cref{eq:znumerical,eq:Laplace-num-property,eq:Qnumerical} together, we find
\begin{subequations}
\begin{align}
    \hat{Z}(s)&=\frac{4\pi\lambda_B}{\varrho_pe}\frac{\varrho_p^2}{D}\frac{1}{\beta e}\hat{Z}_{\rm num}\\
    \hat{Z}_{\rm num}&\equiv\frac{\bar{\Psi}_0}{\im \bar{\omega}}\left[\bar{Q}(t_\text{max})\me^{-\im \bar{\omega} \bar{t}_\text{max}} - \bar{Q}(0)+\im \bar{\omega} \bar{\mathcal{L}}^{\text{num}}\big\{\bar{Q}\big\}\right]^{-1}\,,\label{eq:Znum2}
\end{align}
\end{subequations}
where $\bar{\omega}=\omega \varrho_p^2/D$ and $\bar{\mathcal{L}}^{\text{num}}\left\{\right\}=\mathcal{L}^{\text{num}}\left\{\right\} D/\varrho_p^2$.
Using that $\lambda_B=1/(8\pi c_0 \lambda_D^2)$ and that, in our PNP framework, the electrolyte's conductivity is $\kappa=2 e^2 D\beta c_0$, we find
\begin{equation}
    \hat{Z}(s)=\pi\frac{\ell_p}{\kappa \pi \varrho_p^2}\frac{\varrho_p^3}{\ell_p\lambda_D^2}\hat{Z}_{\rm num}.
\end{equation}

In $\ell_p/(\kappa \pi \varrho_p^2)$, we recognize the resistance $R_p$ of an ideal cylindrical pore filled with a dilute electrolyte [\cref{eq:Rpore}].
Therefore, for thin EDLs, we can compare $\hat{Z}(s)/R_p=\pi\varrho_p^3/(\ell_p\lambda_D^2)\hat{Z}_{\rm num}$ directly to $(Z_p+R_{r})/R_p$ [\cref{eq:ZWO_total}].
For thick EDLs, we will compare $\hat{Z}(s)/R_p=\pi\varrho_p^3/(\ell_p\lambda_D^2)\hat{Z}_{\rm num}$ to \cref{eq:ZfromstepGupta}. 
Note that in numerically performing the Laplace transform in $\hat{Z}_{\rm num}$, we use $Q(t)$ data for many more times than what we plotted in \cref{fig:step_response}.
Moreover, we note that the initial surface charge $\bar{Q}(0)$ in \cref{eq:Znum2} is nonzero.
Physically, one applies a potential difference at $t=0$ between pores by connecting them to a voltage source. 
The time it takes to apply this potential is set by the speed of electric signals in the external wiring. 
Meanwhile, the electric field in our geometry will relax accordingly on the dielectric relaxation time of the solvent, which is orders of magnitude faster than the ionic dynamics.
We thus interpret $\bar{Q}(0)$ as the surface charge after the potential has been applied but before ions have moved.
We determine $\bar{Q}(0)$ of the different pore-reservoir systems by a separate simulation of the Laplace equation---\cref{eq:nondim-numPNP1} with its right-hand side set to zero, subject to \cref{eq:appliedpotential,eq:nonchargedboundary}. 

\subsubsection{Impedance for thin EDLs}
\Cref{fig:impedance} shows the numerically-determined impedances (black lines) for the same parameters as used in \cref{fig:step_response}. 
This figure also shows \cref{eq:ZWO_total} (black dotted lines), with $R_p/R_{r}$ determined similarly to \cref{sec:step_response_numerically}.
The TL model decently approximates the impedance of finite-length pores for aspect ratios beyond $\ell_p/\varrho_p>5$.
For the smaller aspect ratios and at high frequencies, the numerical impedances deviate from the 45-degree phase angle associated with $Z_p$, tending towards a pure capacitance (90 degrees).
Notice that the high-frequency discrepancies nicely correspond to the early-time discrepancies of \cref{fig:step_response}, as high frequencies in EIS correspond to fast processes. 
This means that improved models for $R_{r}$ and $R_p$ cannot fix all the TL model's problems, as changing $R_{r}/R_p$ will merely shift $Z_p$ horizontally and not affect the high-frequency phase angle.
Improved TL models should instead model the early-time nonlinear potential in the reservoir.

\begin{table}
\begin{ruledtabular}
\centering
    \begin{tabular}{cccc} 
    $\ell_p/\varrho_p$ & $Z(\bar{\omega}_{\rm max})/R_p$  & $R_{r}/R_p$ [\cref{eq:idealizedR}]& $R_{r}/R_p$ (impedance.py)\\ 
    \hline
    1   & 1.065  & 0.885  &$\SI{0.948}{}\pm \SI{0.005}{}$  \\ 
    2.5 & 0.441  & 0.354  &$\SI{0.379}{}\pm \SI{0.002}{}$   \\ 
    5   & 0.218  & 0.177  &$\SI{0.184}{}\pm \SI{0.001}{}$  \\ 
    10  & 0.110  & 0.0885  &$\SI{0.0900}{}\pm \SI{0.0004}{}$\\ 
    25  & 0.0450  & 0.0354  &$\SI{0.0359}{}\pm \SI{0.0001}{}$\\ 
    \end{tabular}
    \caption{Values of the high-frequency limit of $Z/R_p$, which, according to \cref{eq:ZWO_total}, should be $R_{r}/R_p$. We present data for $Z(\bar{\omega}_{\rm max})/R_p$, with $\bar{\omega}_{\rm max}=10^4$, from \cref{eq:idealizedR} and a complex nonlinear least square fit of \cref{eq:ZWO_total} to the numerical PNP data using impedance.py \cite{Murbach2020}.}
\label{table:RandRbvalues}
\end{ruledtabular}
\end{table}

We compare the high-frequency limits of the numerical data and analytical predictions in \cref{table:RandRbvalues}.
The second column shows $\Re(Z(\bar{\omega}_{\rm max}))/R_p$ as obtained by PNP, where we used $\bar{\omega}_\text{max}=10^4$, at which point $\Im(Z)$ is negligible.
The third column lists the high-frequency limit of $(R_{r}+Z_p)/R_p$, that is, $R_{r}/R_p$, which we determined for the respective parameters by \cref{eq:idealizedR}.
In line with our observations of \cref{fig:impedance}, deviations between these two methods are larger for smaller aspect ratios.
Next, we performed complex nonlinear least square fits of $(R_{r}+Z_p)/R_p$ [\cref{eq:ZWO_total}] to the numerical PNP data using impedance.py \cite{Murbach2020}, with $R_{r}/R_p$ and $R_p C $ as fit parameters.
Representative fits are shown for $\ell_p/\varrho_p=1$ and 2.5 with purple dashed lines.
We also performed fits for all other aspect ratios, for which we list the fit parameter $R_{r}/R_p$ in the last column of \cref{table:RandRbvalues}.
Even for the large aspect ratio $\ell_p/\varrho_p=25$, the numerical data and the $R_{r}/R_p$ fit parameter differ substantially.
Hence, even for the system for which the TL model was devised---a long pore subject to a small potential, in contact with an electrolyte reservoir filled with dilute electrolyte---there is no one-to-one relation between the TL model's fit parameters $R_{r}/R_p$ and $R_p C$ on the one hand and the microscopic parameters characterizing the pore geometry and electrolyte properties on the other.

\begin{figure}
\includegraphics[width=\linewidth]{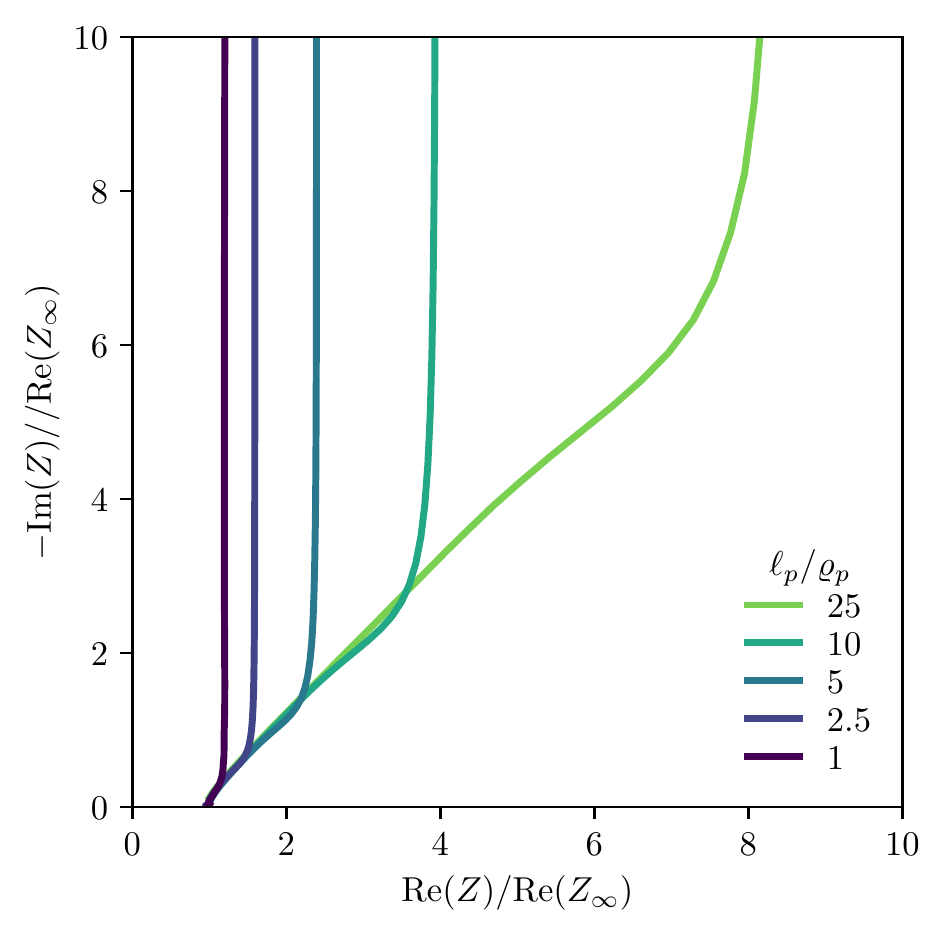}
    \caption{Impedances $Z$ of pores of different lengths, determined from PNP step voltage solutions. This is based on the same data as the black lines in \cref{fig:impedance}, but now all curves are scaled to their high-frequency limit, $\Re(Z_{\infty})=\Re(Z(\omega_{\rm max}))$.}
    \label{fig:impedance_replotted}
\end{figure}

\Cref{fig:impedance_replotted} shows the same PNP data for $Z$ as in \cref{fig:impedance} but now scaled to $\Re(Z(\omega\to\infty))$ instead of $R_p$.
This data representation corresponds more clearly to experiments on porous electrodes of various widths \cite{lust2004influenceII, eikerling2005optimized,kotz2000principles,ogihara2015impedance}.
Moreover, this data representation shows that decreasing the pore length leads to a smaller pore resistance; in the TL model, the pore's resistance is set by the difference between the high and low-frequency limits of the impedance, $R_p=3\{\Re[Z(\omega\to0)]-\Re[Z(\omega\to\infty)]\}$.
We conclude that decreasing pore length leads to impedance curves that progressively move towards that of a pure capacitor, as 1) the 45-degree line becomes shorter, and 2) the high-frequency regime deviates from 45 degrees (clearer visible in \cref{fig:impedance}).

\begin{figure*}
\includegraphics[width=\linewidth]{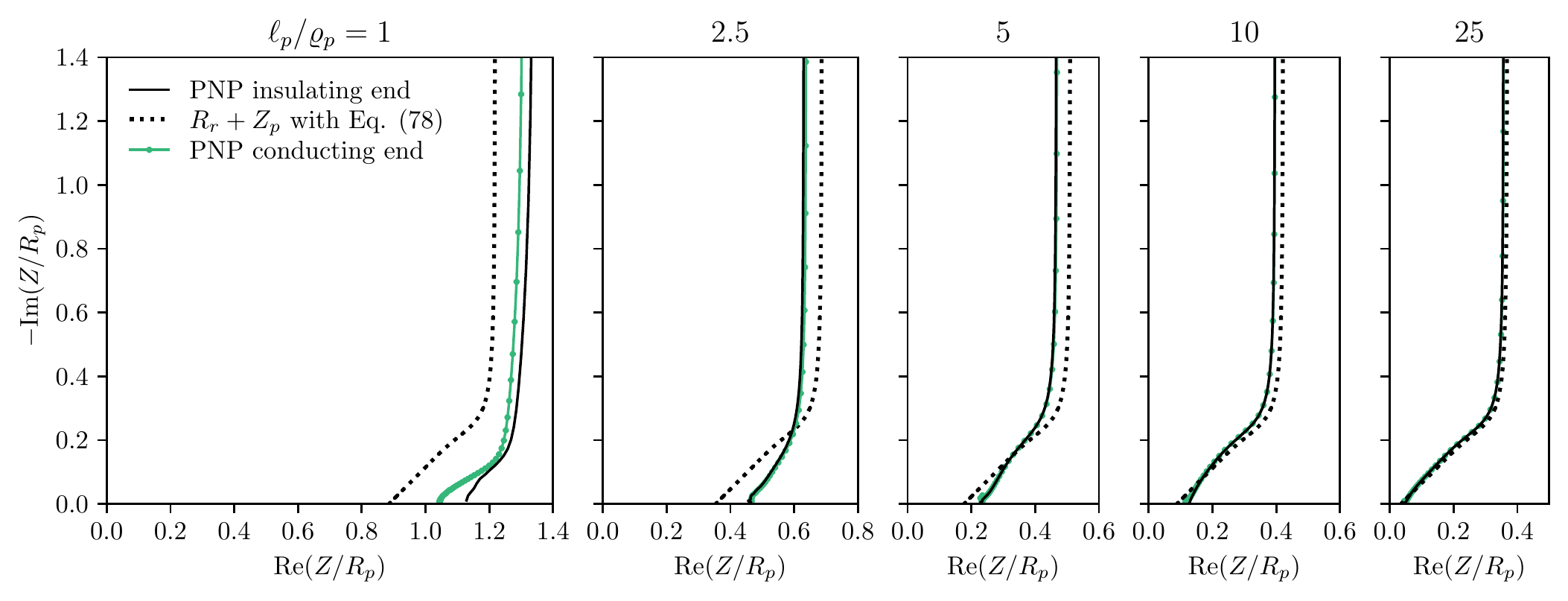}
    \caption{
    Impedances $Z$ of pores of different aspect ratios $\ell_p/\varrho_p$ with overlapping EDLs ($\lambda_D/\varrho_p=1$) and other parameters as in \cref{fig:step_response_Henrique}.
    The data corresponds to \cref{eq:ZfromstepGupta} (dotted) and PNP step voltage solutions for insulating (black lines) and conducting (green lines with circles) pore ends.
    The impedance \cref{eq:ZfromstepGupta} is scaled to the pore resistance $R_p$ [into which we absorbed an EDL overlap dependent prefactor, see below \cref{eq:ZfromstepGupta}], and the PNP data is scaled to $\ell_p/(\kappa\pi \varrho_p^2)$, i.e., the resistance of a cylindrical pore with an electrolyte at infinite dilution.}
    \label{fig:impedance_thickEDLs}
\end{figure*}

\subsubsection{Impedance for thick EDLs}
\Cref{fig:impedance_thickEDLs} shows the numerically-determined impedances (black lines) and \cref{eq:ZfromstepGupta} (black dotted lines) for the same parameters as in \cref{fig:step_response_Henrique}. 
Again, the theoretical prediction performs decently for large aspect ratios but not for smaller ones.
Overall, taking $\varrho_p/\ell_p=5$ as an example, the fit between numerics and theory is better in \cref{fig:impedance} than in \cref{fig:impedance_thickEDLs}.
In our discussion of \cref{fig:step_response_Henrique}, we noted that EDL overlap leads to more involved centerline potentials than in the nonoverlapping case (\cref{fig:step_response}): $\psi_c(z,t)$ transitions at the reservoir-pore interface from a small value in the reservoir to a finite value in the pore, even at late times.
Henrique, Zuk, and Gupta's model captured the late-time in-pore centerline potential well.
Conversely, the transition at the pore-reservoir interface was not captured, and the late-time centerline potential of short pores was overestimated.
These two points may have led to the larger discrepancies between numerics and theory for short pores in \cref{fig:impedance_thickEDLs} than in \cref{fig:impedance}.

\subsubsection{Impedance of pores with conducting ends}\label{sec:conductingends}
So far, we discussed pores whose cylindrical surface was conducting but whose ends ($\Gamma_1$ and $\Gamma_7$ in \cref{fig:schematic}) were insulating.
That boundary condition corresponds to the experiments of Eloot and coworkers \cite{eloot1995calculationII} on pores drilled into stainless steel and insulating plexiglass at their ends.
Conversely, pores in supercapacitor have conducting carbon surfaces on all sides except their opening.
To describe such pores, we change \cref{eq:numBCs_potential,eq:numBCs_nocharge} to  
\begin{subequations}
\begin{align}
    -\psi\big |_{\Gamma_1}=-\psi\big |_{\Gamma_2} = \psi\big |_{\Gamma_6}= \psi\big |_{\Gamma_7} &= \Psi_0,\\
    \boldsymbol{\nabla}\psi\cdot\boldsymbol{n}\big |_{\Gamma_3, \Gamma_4,\Gamma_5,\Gamma_8} &= 0. 
\end{align}
\end{subequations}
As a result, \cref{eq:appliedpotential,eq:nonchargedboundary} change to
\begin{subequations}
\begin{align}
    -\bar{\psi}\big |_{\Gamma_1} =-\bar{\psi}\big |_{\Gamma_2} = \bar{\psi}\big |_{\Gamma_6}=\bar{\psi}\big |_{\Gamma_8} &= \bar{\Psi}_0,\\
    \partial_{\bar{\rho}}\bar{\psi} |_{\Gamma_4,\Gamma_8}=\partial_{\bar{z}}\bar{\psi} |_{\Gamma_3,\Gamma_5}&=0,
\end{align}
\end{subequations}
\Cref{eq:variational1} changes to  
\begin{align}       
&-\int_{\Gamma_1}\partial_{\bar{z}}\bar{\psi} v \bar{\rho}\diff \bar{\rho} 
+\int_{\Gamma_2,\Gamma_6}\partial_{\bar{\rho}}\bar{\psi} v \diff \bar{z}
+\int_{\Gamma_7}\partial_{\bar{z}}\bar{\psi} v \bar{\rho}\diff \bar{\rho} \nn 
&- \int_{\Omega}\partial_{\bar{\rho}}\bar{\psi}\partial_{\bar{\rho}} v \bar{\rho} \diff \bar{\rho}\diff \bar{z} -\int_{\Omega}\partial_{\bar{z}}\bar{\psi}\partial_{\bar{z}} v \bar{\rho}\diff \bar{\rho}\diff \bar{z} \nn
    &\qquad=-\frac{1}{2}\frac{\varrho_p^2}{\lambda_D ^2}\int_{\Omega}(\bar{c}_+ - {\bar{c}_-})v\bar{\rho}\diff \bar{\rho}\diff \bar{z}, 
\end{align}
\Cref{eq:surfacechargedensity} changes to
\begin{equation}
    \sigma = \frac{1}{4\pi\lambda_B}
    \begin{cases}
        \partial_\rho \bar{\psi}\quad &\textrm{on}\quad\Gamma_6\,,\\
        \partial_z \bar{\psi}\quad &\textrm{on}\quad\Gamma_7\,,
    \end{cases}
\end{equation}
and \cref{eq:Qnumerical} for the total charge on one pore $Q=e\int_{\Gamma_6,\Gamma_7}\diff A\, \sigma$ becomes
\begin{align}
    Q&=\frac{\varrho_p e}{2\lambda_B}\left(\int_{0}^{\ell_p/\varrho_p}\diff \bar{z} \,\partial_{\bar{\rho}} \bar{\psi}(1,\bar{z},\bar{t})\right.\nn
    &\qquad\qquad\qquad\left.+\int_{0}^{1}\diff \bar{\rho} \,\bar{\rho} \partial_{\bar{z}} \bar{\psi}(\bar{\rho},\ell_p/\varrho_p,\bar{t})\right)\,.
\end{align}

\Cref{fig:impedance,fig:impedance_thickEDLs} show the impedance of pores with conducting ends of various lengths (green lines with circles) as obtained from PNP solutions. 
For $\ell_p/\varrho_p=10$ and $25$, these data hardly differ from the impedance of pores with insulating ends. 
For shorter pores, differences between both boundary conditions appear,  which makes sense as a relatively larger part of the pore's charged surface area comes from its end.  
For short pores and thin EDLs [\cref{fig:impedance}], the data differ  mainly at low frequencies; for thick EDLs [\cref{fig:impedance_thickEDLs}], they differ mainly at high frequencies.

\section{Discussion}\label{sec:discussion}
\subsection{The term ``Diffusion impedance''}
In the context of pore charging through EDL formation at blocking electrodes, the commonly-used terminology ``diffusion impedance'' is a misnomer \cite{huang2018diffusion}. 
As we showed in \cref{fig:potential} (see also Fig.~(3) of Henrique, Zuk, and Gupta \cite{henrique2021charging} \footnote{Even though it looks similar to \cref{fig:potential}, note that Fig.~(3) of \cit{henrique2021charging} is not to scale, and corresponds to aspect ratios between $\ell_p/\varrho_p=10$ and 50 (\emph{private communication with F. Henrique})}), ions flow into a pore by electromigration, not diffusion.
Dropping the all-important electromigration terms in the PNP equations yields a regular ionic diffusion equation. 
Hence, solving the ionic diffusion equation to find an electrode's impedance, as was done, for example, in \cit{2017cooper}, does not account for the relevant physics (as these authors acknowledged).
Nevertheless, \cit{2017cooper} found sensible impedances from the perturbed ion densities.
How can this be?
We have seen in this article that de Levie's transmission line model, a diffusion-type equation for the potential drop $\psi_d$, accurately describes the relaxation of a pore's centerline potential.
Solving a diffusion equation for ionic species and determining the impedance from the perturbed densities yields the correct impedance, as the mathematical form of all the equations is the same as the ones we used to derive the pore impedance $Z_p$ from the TL equation (but in 3d). 
Hence, \cit{2017cooper} solved the correct diffusion-type equation, but the diffusing quantity is the centerline potential $\psi_d$, not the ions.

\subsection{Towards porous electrodes: $m$ parallel pores vs. stack electrode model}
So far, we have discussed charging a single cylindrical pore in contact with a large reservoir. 
Different models were proposed to go from known single-pore charging behavior to predict the charging of a complete porous electrode. 
Here, we compare two models for an electrode with $m$ pores.

\begin{figure}
    \includegraphics[width=\linewidth]{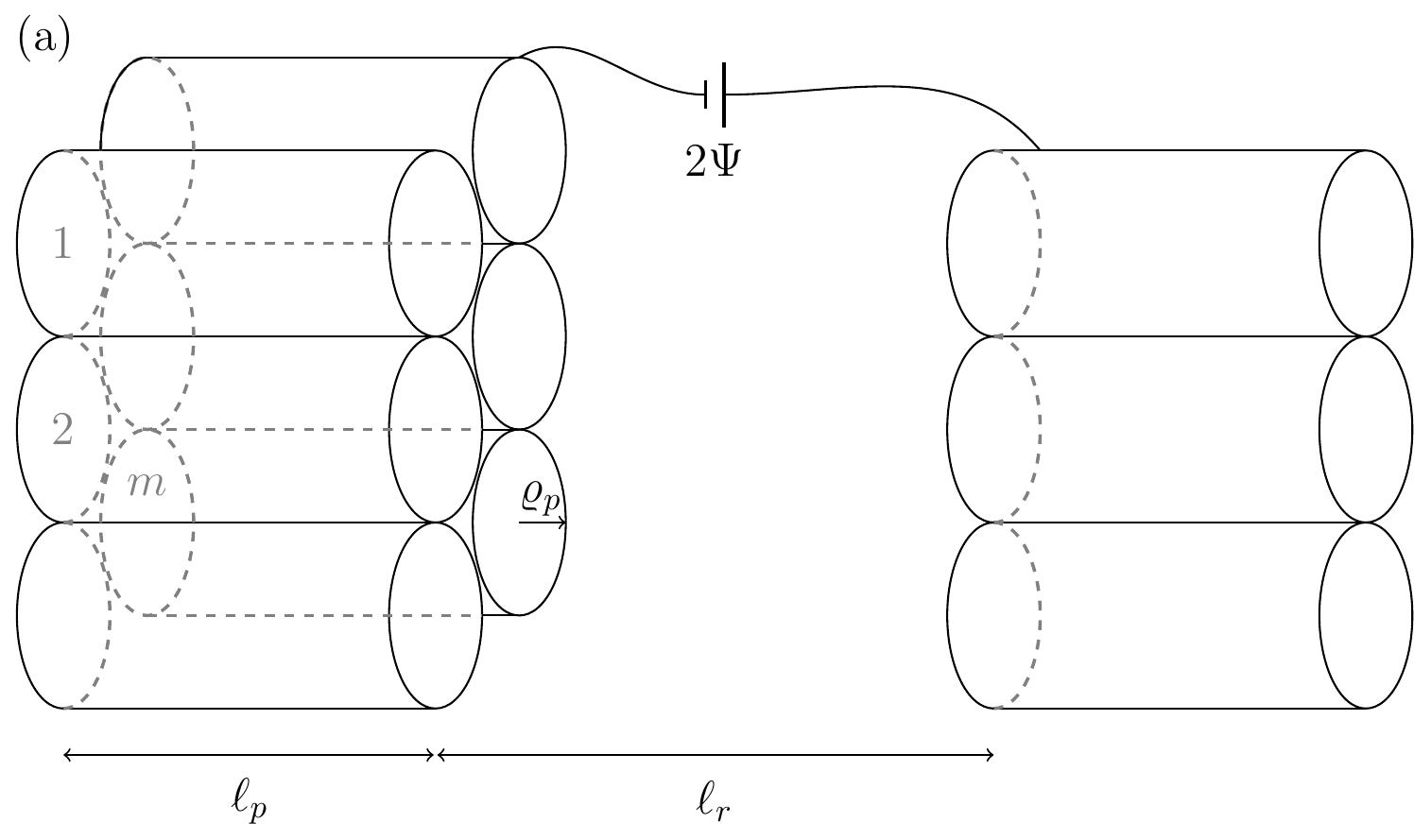}
    \includegraphics[width=\linewidth]{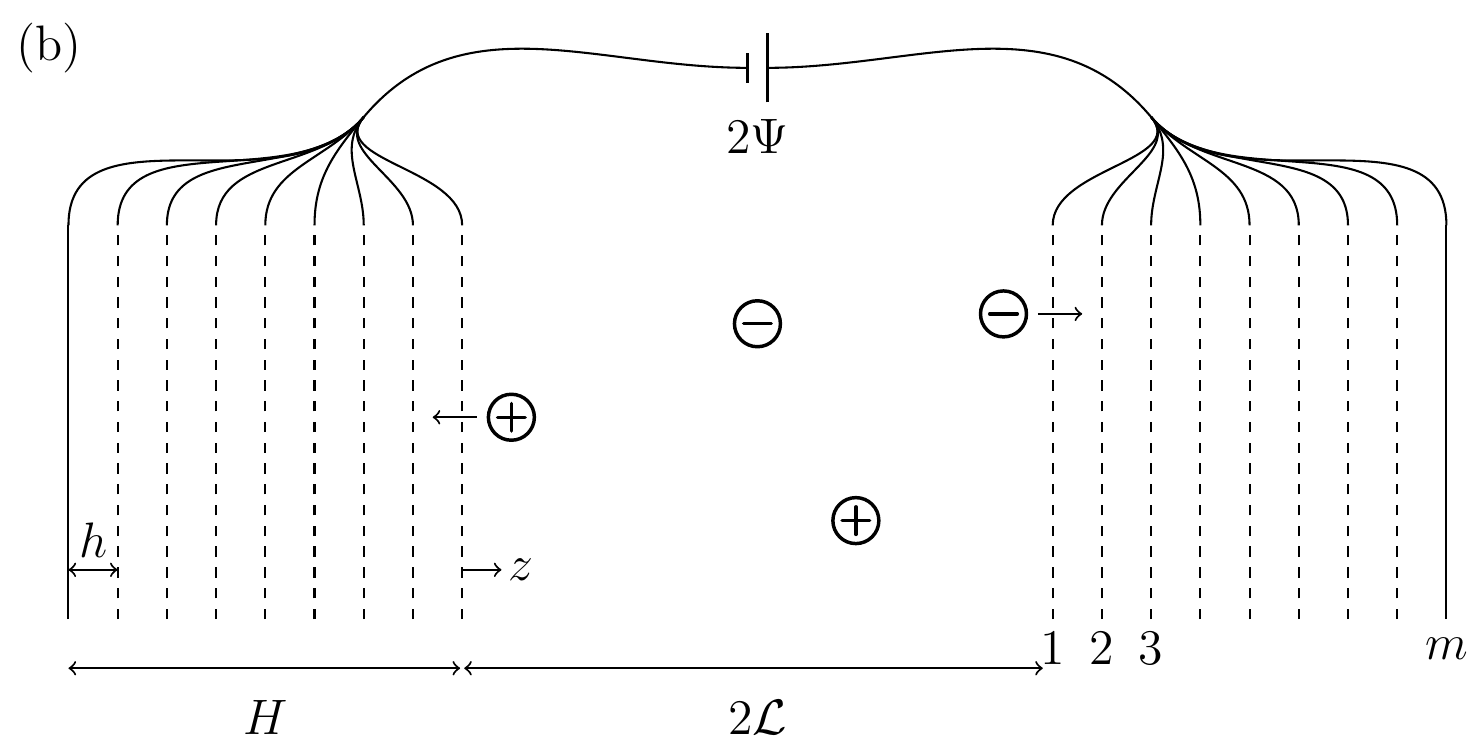}
    \caption{$m$ parallel pores (a) and stack electrode model (b). \label{fig_parallel_pores}}
\end{figure}

Several papers treated porous electrodes as a bundle of $m$ cylindrical pores connected in parallel  \cite{barcia2002application,lasia2014electrochemical,cericola2016impedance}; see \cref{fig_parallel_pores}(a).
In this case, the impedance of both electrodes and reservoir amounts to
\begin{equation}\label{eq:Warburgnpores}
    Z_\text{tot}=R_{r}+\frac{2}{m}Z_p\,.
\end{equation}
The current in response to a step potential, for which $\hat{\Psi}(s)=\Psi_0/s$, is then $I(t)=\Psi_0\mathcal{L}^{-1}\left\{1/(s Z_\text{tot})\right\}$.
The relaxation time of this system is set by the zeros of $Z_\text{tot}$, that is, by the solution to
\begin{equation}
    \frac{\coth\sqrt{sR_p C }}{\sqrt{sR_p C }}+\frac{mR_{r}}{2R_p}=0\,.
\end{equation}
Substituting $sR_p C =-\alpha_j^2$ gives
\begin{equation}
    \alpha_j \tan\alpha_j=\frac{2R_p}{mR_{r}}\,,
\end{equation}
which, up to the factor $m$, is the same as in Janssen \cite{janssen2021transmission} [and \cref{eq:continuumtranscendental} here].
An approximate solution based on Pad\'{e} approximation reads $\alpha_j^{-1}\approx\sqrt{1/3+mR_{r}/(2R_p)}$, which yields the relaxation time 
\begin{equation}\label{eq:nparallelporetau}
    \tau=\frac{R_p C }{\alpha_j^2}\approx\frac{1}{3}R_p C +\frac{m}{2} R_{r} C\,.
\end{equation}
With $R_p=\ell_p/(\kappa \pi \varrho_p^2)$,  $C=\varepsilon 2\pi \varrho_p \ell_p/\lambda_D$, and $\kappa=\varepsilon D/\lambda_D^2$ we find $R_p C =2\lambda_D \ell_p^2/(D\varrho_p)$.
To express the reservoir resistance $R_{r}=\ell_r/(\kappa A^c)$, 
we equate the reservoir's cross-sectional area $A^c$ (perpendicular to the pores) to that of the pore-bundle electrode.
Assuming no space to be left between the pores, each having a radius $\varrho_p$, yields $A^c=m \pi\varrho_p^2$. 
Collecting terms, we find
\begin{equation}\label{eq:nparalleltau}
    \tau=\frac{\lambda_D \ell_p^2}{D\varrho_p}\left(\frac{2}{3}+ \frac{\ell_r}{ \ell_p}\right)\,,
\end{equation}
which, notably, does not depend on $m$.

Lian and coworkers~\cite{lian2020blessing} recently proposed an alternative model for porous electrode charging.
In their ``stack electrode'' model [\cref{fig_parallel_pores}(b)], the two porous electrodes of a supercapacitor, separated by $2\mathcal{L}$ and both of width $H$, are represented by $m$ flat electrode ``sheets'' spaced $h$ apart [so that $H=h(m-1)$] \cite{lian2020blessing,lin2022microscopic,ji2023asymptotic}.
Of these sheets, the outer ones are blocking, while the others are fully permeable to ions. 
Upon applying a potential difference to the two porous electrodes, with each sheet in an electrode at the same potential, ions move perpendicular to the sheets and through them, forming EDLs on both sides of each sheet (except the outer sheets).
When the lateral size of the sheets is much larger than the width $2H+2\mathcal{L}$ of the setup, the potential and ion densities depend only on the coordinate $z$ perpendicular to the sheets.
Lian and coworkers~\cite{lian2020blessing} solved the PNP equation in this effectively one-dimensional geometry to determine each sheet's time-dependent surface charge.
They showed that the stack electrode model relaxes, for small applied potential, with the same timescale as a discrete TL circuit (\cref{TLcircuit_standard})  with $m$ rungs, with $r_i=r $ and $c_i=2c$ for $i=1,\ldots,m-1$ and $c_m=c$, and total resistance $R=\sum_i r_i$ and capacitance $C=\sum_i c_i$.
For $m\gg1$, this circuit relaxes with almost the same timescale as the regular finite-$m$ TL circuit, whose timescale reads \cite{janssen2021transmission}
\begin{equation}
    \tau\approx \frac{1}{3}R C +R_{r} C .
\end{equation}
Using $R/R_{r}=\mathcal{L}/H$, $C=c(2m-1)$, $R=r(m-1)$, $c=2\varepsilon A/\lambda_D$, $r= h/(\kappa A)$, and $\kappa=\varepsilon D/\lambda_D^2$, one finds that the stack electrode model relaxes on a timescale 
\begin{equation}\label{eq:stackelectrode}
    \tau =(2m-1)\frac{\lambda_D \mathcal{L}}{D}\left(1+\frac{H}{3\mathcal{L}}\right)\,.
\end{equation}

Comparing the two models in \cref{fig_parallel_pores} and identifying $\mathcal{L}\to\ell_r/2$, $H\to\ell_p$, and $h\to \varrho_p$, \cref{eq:nparalleltau} becomes
\begin{equation}
    \tau =2(m-1)\frac{\lambda_D \mathcal{L}}{D}\left(1+\frac{H}{3\mathcal{L}}\right),
\end{equation}
obviously, with differences to \cref{eq:stackelectrode} being subleading in $m$.
For a stack electrode model whose last plate is permeable as well, both models have identical charging times.

The $m$ parallel pores and stack electrode models both utilize the TL circuit, but they do so differently.
The $m$ parallel pores model uses $Z_p$, which we found from the TL circuit in the $n\to\infty$ limit.
In other words, the $m$ parallel pore model uses the $n\to\infty$ circuit $m$ times.
By contrast, $m$ is kept finite in the stack electrode model, with no corresponding $n\to\infty$ limit. 

While the relaxation times of both models are thus the same, their impedances are not, as we saw by comparing \cref{eq:Warburgnpores} to $R_{r}+2Z_1$, with $Z_1$ from \cref{eq:Zn}.
This is unsurprising as the parameter $m$ plays different roles in both models. 
In the $m$ parallel pore model, increasing $m$ corresponds to using electrodes with a larger cross-sectional area.
The stack electrode model, by contrast, is one dimensional, so it models a porous electrode per unit cross-sectional area.
Increasing $m$ in the stack electrode model corresponds to using thicker electrodes (if the pore width $h$ is kept fixed) or using narrower pores (if the electrode thickness $H$ is kept fixed).

\section{Conclusions}\label{sec:conclusion}
We derived the pore impedance $Z_p$ directly from its corresponding TL circuit---to our knowledge, side-stepping the TL equation or other diffusion-type PDEs for the first time.
As the TL circuit and its extension find use in interpreting various electrochemical devices such as batteries and fuel cells \cite{nielsen2014impedance,moskon2021transmission,vivier2022impedance}, our methods could be useful more broadly than for the example of EDL capacitors with porous electrodes that we focussed on here. 
Future work could generalize our calculations to determine the impedance of a groove \cite{delevie1965}, an arbitrarily-shaped pore \cite{keiser1976abschatzung}, or to find the impedance of a case with finite electrode resistance \cite{paasch1993theory}. 

There are at least four lengthscales relevant to the charging of a cylindrical pore: its length $\ell_p$ and radius $\varrho_p$, the width $\lambda_D$ of the EDL, and the combination $\sqrt{D/\omega}$ of the ionic diffusion constant to the angular frequency of the harmonic voltage source. 
Two of the three independent dimensionless combinations of these lengthscales had been characterized.
De Levie showed that a dimensionless penetration depth $\propto \sqrt{D\varrho_p/(\omega \ell_p^2 \lambda_D)}$ sets the characteristic length until ionic density profiles in a pore are perturbed \cite{levie1963};
Henrique, Zuk, and Gupta studied the effect of the EDL overlap $\varrho_p/\lambda_D$ on pore charging.
This left one dimensionless ratio, the pore aspect ratio $\ell_p/\varrho_p$, which had received little attention. 
Accordingly, we studied the charging of pores of various aspect ratios by numerical simulations of the Poisson-Nernst-Planck (PNP) equations.
We found impedances of long pores to agree well with $Z_p$.
By contrast, deviations were visible at high frequencies for pores with aspect ratios less than $\ell_p/\varrho_p=5$.
Our findings are thus in qualitative agreement with Eloot and coworkers~\cite{eloot1995calculationII}, who found that their experimental pore impedance data could not be fitted by equivalent circuits when  $\ell_p/\varrho_p<2$.

The shapes of the impedance curves that we found are not unique to short pores; similar curves resulted, for instance, from an equivalent circuit model accounting for the outer surface of a porous electrode through a parallel connection of $Z_p$ and another capacitor \cite{jurczakowski2004impedance}.
Figure~20 of that article contains experimental impedance data for a porous gold electrode; the shape of their impedance is very similar to ours in \cref{fig:impedance} for $\ell_p/\varrho_p=2.5$.
Hence, above-45 degrees high-frequency phase angles may be explained by at least two distinct phenomena: pore aspect ratio or outer surface capacitance. 
Deciding which applies would require further impedance spectroscopy on different electrodes or different experiments.

We see the following directions for future work.
First, an outstanding challenge is to analytically solve the PNP equations we solved numerically in \cref{sec:numericalstudy}. 
In previous work, we analytically solved the PNP equation for a long pore and negligible reservoir resistance \cite{aslyamov2022analytical}.
Relaxing these restrictions to describe a short pore next to a nonnegligible reservoir will be challenging. 
Second, the boundary conditions of the PNP equations can be adapted to pores with curved \cite{keiser1976abschatzung}, rough \cite{delevie1965,gunning1995exact,delevie1990fractals,Aslyamov2022Properties,aslyamov2021electrolyte,seebeck2022elucidating}, or nonblocking surfaces \cite{biesheuvel2011diffuse,li2022impedance}.
Third, the PNP model should be extended with finite ion sizes and dispersion and image charge interactions when pores are very narrow  \cite{kondrat_accelerating_2014,tomlin2022impedance} or large potentials are applied, for instance, when probing a system's nonlinear impedance \cite{Kirk_2023, hallemans2023electrochemical} 
or its impedance around a large bias voltage.
Large applied potentials cause diffusive salt transport not captured by the TL model \cite{aslyamov2022analytical}, so a pore's impedance will deviate from $Z_p$.
Last, this article aimed at bringing equivalent circuit and continuum modeling of electrolyte-filled pores closer together. 
It would be interesting to do the same for the equivalent circuit models and molecular dynamics simulations \cite{pean2014dynamics,bi2020molecular,jeanmairet2022microscopic}; that is, to pinpoint the meaning of fit parameters when the TL model is fitted to molecular dynamics data.

\begin{acknowledgements}
We thank Filipe~Henrique for giving detailed comments on our manuscript.
\end{acknowledgements}

\end{document}